\documentclass[aps,10pt,notitlepage,twocolumn,nofootinbib,superscriptaddress]{revtex4-2}

\usepackage{graphicx,color}
\usepackage{amsmath}
\usepackage{amssymb}
\usepackage{hyperref}
\usepackage[utf8]{inputenc}
\usepackage[english]{babel}
\usepackage{epsfig}
\usepackage{subfigure}
\usepackage{wasysym}
\usepackage{color,xcolor}
\usepackage{amsmath}
\usepackage{bm}
\usepackage{epsfig}
\usepackage{amsfonts}
\usepackage{dcolumn}
\usepackage{float}
\usepackage{booktabs}
\usepackage{relsize}

\hypersetup{colorlinks=true, linkcolor=blue, citecolor=green}

\usepackage{dcolumn}% Align table columns on decimal point
\usepackage{bm}% bold math
\usepackage{ifpdf}
\usepackage{hyperref}
\usepackage{float}
\usepackage{bm}
\usepackage{xcolor,color,graphicx,graphics}
\usepackage[OT1]{fontenc}
\usepackage{latexsym,amssymb,amsmath,amsfonts}
\usepackage{makeidx}
\usepackage{epsfig}
\usepackage{epstopdf}
\usepackage{mathrsfs}
\hypersetup{colorlinks=true, linkcolor=blue, citecolor=green}
\usepackage{enumerate}
\usepackage{xcolor}
 \usepackage{multirow}
\begin{document}
%========================================================

\title{Supermassive black hole in NGC 4649 (M60) with a dark matter halo:\\ Impact on shadow measurements and thermodynamic properties}

	\author{Francisco S. N. Lobo} \email{fslobo@ciencias.ulisboa.pt}
\affiliation{Institute of Astrophysics and Space Sciences, Faculty of Sciences, University of Lisbon, Building C8, Campo Grande, P-1749-016 Lisbon, Portugal}
\affiliation{Department of Physics, Faculty of Sciences, University of Lisbon, Building C8, Campo Grande, P-1749-016 Lisbon, Portugal}

     \author{Jorde A. A. Ramos}
    \email{jordealves@ufpa.br}
\affiliation{Faculty of Physics, Graduate Program in Physics, Federal University of Pará, 66075-110, Belém, Pará, Brazil}

    \author{Manuel E. Rodrigues} \email{esialg@gmail.com}
\affiliation{Faculty of Physics, Graduate Program in Physics, Federal University of Pará, 66075-110, Belém, Pará, Brazil}
\affiliation{Faculty of Exact Sciences and Technology, Federal University of Pará, Abaetetuba University Campus, 68440-000, Abaetetuba, Pará, Brazil}

%%%%%%%%%%%%%%%%%%%%%%%%%%%%%%%%%%%%%%%%%%%%%%%%%%%%%%%%%%%%%%%%

%%%%%%%%%%%%%%%%%%%%%%%%%%%%%%%%%%%%%%%%%%%%%%%%%%%%%%%%%%%%%%%%
%========================================================
\begin{abstract}
%========================================================

We investigate black hole (BH) solutions embedded in a dark matter (DM) halo, modeled as extensions of the Schwarzschild metric. The DM density profile is constrained by Hubble Space Telescope (HST) data, stellar dynamics, and globular cluster (GC) measurements of the elliptical galaxy NGC 4649 (M60). Using this profile, we construct two distinct spacetime solutions characterized by the black hole mass ($M$), critical velocity ($V_c$), and core radius ($a$), all reducing to the Schwarzschild case in the limit $V_c = 0$ and $a = 0$.
Our results show that the DM halo modifies essential BH features, such as the event horizon radius and spacetime curvature, as reflected in the Kretschmann scalar. We also derive an approximate analytical expression for the BH shadow radius, which increases slightly due to the halo’s influence. Comparisons with two observational datasets further validate the analysis.
Thermodynamic properties are examined across the two models. In the first, a generalized Smarr formula is obtained via two additional variables, $\gamma$ and $a$. The second incorporates halo effects through $V_c$ and $a$. These results underscore the role of DM in shaping both geometric and thermodynamic aspects of BHs.

\end{abstract}
%========================================================
\pacs{04.50.Kd,04.70.Bw}
%=================================================================
\date{\today}
%========================================================
\maketitle
%=================================================================
\def\HMS{{\scriptscriptstyle{\rm HMS}}}
%========================================================
%\bigskip
%\hrule
%\tableofcontents
%\bigskip
%\hrule
%========================================================
%\parindent0pt
%\parskip7pt
%========================================================

%%%%%%%%%%%%%%%%%%%%%%%%%%%%%%%%%%%%%%%%%%%%%%%%%%%%%%%%%%%%%%%%
\section{Introduction}\label{sec1}
%%%%%%%%%%%%%%%%%%%%%%%%%%%%%%%%%%%%%%%%%%%%%%%%%%%%%%%%%%%%%%%%

The Schwarzschild solution \cite{schw1999} to Einstein’s field equations, formulated within the framework of general relativity \cite{Wald-RG1984}, describes a non-rotating, spherically symmetric compact object of infinite density—now recognized as a black hole (BH). These entities are enveloped by an event horizon, a causal boundary beyond which no information can escape, and they exhibit intense gravitational phenomena that have profoundly influenced our conceptualization of spacetime and matter. 
Once believed to trap all matter and radiation, BHs were later shown to emit thermal radiation due to quantum effects, as demonstrated by Hawking \cite{S-Hawking1975}. This groundbreaking discovery led to the development of black hole thermodynamics \cite{gibbons1977, CT-Davies1977, toussaint1979, hawking1983, york1986, gibbons1996}, which revolutionized the theoretical landscape by bridging general relativity, quantum field theory, and thermodynamic principles. In contemporary physics, BHs serve as critical arenas for probing the unification of fundamental interactions under extreme conditions.

Another indispensable component of the cosmos is dark matter (DM), the existence of which is now firmly established through various lines of observational evidence \cite{bertone2005, KFreese2008, Swart2017, Risa2018, Arbey2021}. Contributing approximately five-sixths of the universe's total matter content \cite{Planck2021}, DM interacts primarily through gravitation, as no conclusive detection of its non-gravitational interactions has yet been achieved \cite{cebrian2022, Misiaszek2023}. In cosmological models, DM plays a central role in driving the formation and evolution of large-scale structures. Galaxies are thought to coalesce within extensive DM halos, which shape their rotation curves and mass distributions \cite{Valluri2003}. Understanding the nature and dynamics of DM is thus critical for advancing modern astrophysics and cosmology.

The quest to identify plausible DM candidates has led to several extensions of the Standard Model of particle physics \cite{bertone2005, axions}. Such candidates must be electrically neutral, colorless, and either absolutely stable or possess a lifetime exceeding the current age of the universe, $t_{\mathrm{DM}} > t_u \simeq 4.3 \times 10^{17}$ s \cite{bertone2005, axions}. These particles must also comply with cosmological constraints, including those from the cosmic microwave background and large-scale structure surveys \cite{bertone2005, axions, BaixaMassa-AR, HistoriaDM}. Among the most studied candidates are axions \cite{bertone2005, BaixaMassa-AR, HistoriaDM}, which arise from Peccei–Quinn symmetry, as well as weakly interacting massive particles (WIMPs) such as neutralinos \cite{bertone2005, Feng2010}. Other theoretical constructs include superWIMPs like gravitinos \cite{bertone2005, Feng2010}, axinos \cite{bertone2005, axions}, and Kaluza–Klein particles predicted by models with extra spatial dimensions \cite{klein1926}, particularly in the context of universal extra dimensions \cite{UED}. A more thorough overview of these scenarios is available in \cite{bertone2005, BaixaMassa-AR}.

Supermassive black holes (SMBHs), often found at the centers of galaxies \cite{SagittariusA2023, M87SBH2019}, are frequently studied in conjunction with surrounding DM halos. Their mutual interactions are a subject of considerable interest, with some theoretical models predicting the ejection of SMBHs due to dynamical instabilities in dense environments \cite{ElZant2020}. Alternative hypotheses suggest that scalar field DM with attractive self-interactions could facilitate SMBH formation by modifying the core–halo mass relationship \cite{Padilla2021}. Observational efforts have also sought to correlate the mass of central black holes (CBHs) with the surrounding DM halo and stellar remnants, as a means to understand galaxy formation and co-evolutionary processes \cite{Marasco2021, Powell2022, Bansal2023}. These investigations underscore the importance of considering both baryonic and non-baryonic matter in galactic dynamics.

The spatial distribution of DM in halos is typically described using empirical density profiles that capture the radial dependence of the mass concentration \cite{TBSW1997-perfisDM}. One well-known model is the Hernquist profile \cite{Hernquist1990}, given by $\rho(r) \propto \frac{\alpha}{r} \frac{1}{(r+\alpha)^3}$, where the parameter $\alpha$ determines the halo's central concentration \cite{TBSW1997-perfisDM, Liu2022, Shen2024, Fard2024}. Another widely employed model is the Navarro–Frenk–White (NFW) profile \cite{NFW1997}, which follows $\rho(r) \propto \left(\frac{r}{\beta}\right)^{-1} \left(1 + \frac{r}{\beta}\right)^{-2}$ and uses $\beta$ as a scale radius \cite{TBSW1997-perfisDM, AA01, Tsuchiya2013}. Supported by results from $N$-body simulations, the NFW profile predicts a central density slope of $\rho \propto r^{-1}$ \cite{Lu2006}. Despite its success, the model remains under scrutiny due to its inconsistency with certain observational data.

Near SMBHs, DM densities can be amplified significantly, leading to the formation of sharp central \textit{spikes} \cite{Gondolo1999, Sadeghian2013}. This behavior is in contrast with the \textit{cuspy} profiles predicted by models like NFW, which do not align well with observed core-like structures in dwarf galaxies and low surface brightness systems \cite{deblok2010}. To reconcile these discrepancies, modified profiles have been proposed. Notably, a modified NFW profile tailored for self-interacting DM was introduced in \cite{Tran2024}, yielding a flatter core that better fits observational data. However, this approach does not fully capture the early-time physics associated with core formation, highlighting ongoing theoretical limitations in modeling the inner regions of DM halos.

The early universe likely provided conditions favorable for SMBH formation via DM-driven processes \cite{Balberg2002, BSI2002}. Nonetheless, the microphysical mechanisms governing DM behavior in the vicinity of black holes remain poorly understood, complicating attempts to derive comprehensive analytic solutions. Addressing this gap, \cite{AA01} investigated the stationary configuration of BHs surrounded by DM. Their analysis showed modest deviations from the standard Kerr geometry, such as a slight expansion of the event horizon—on the order of $10^{-7}$—as well as a reduction in the size of the ergosphere, while the singularity remained unaffected. These results emphasize the potential subtle but measurable effects of DM on BH spacetimes.

A more elaborate mass distribution model was presented in \cite{mod1} for the massive elliptical galaxy NGC 4649 (M60). Utilizing axisymmetric orbit superposition methods and data from the Hubble Space Telescope (HST), stellar kinematics, and globular cluster (GC) dynamics, the authors incorporated a DM halo with density $\rho_{\mathrm{DM}}$, in conjunction with stellar mass and a central BH. Key physical parameters—including the black hole mass $M$, the core radius $a$, and the critical scalar velocity $V_c$—were inferred. This comprehensive modeling framework allows for a more accurate depiction of the gravitational potential and matter distribution in massive galaxies, and serves as a reference for investigating similar astrophysical systems.

A particularly fascinating observable of BHs is their \textit{shadow}, a dark silhouette delineated by a bright photon ring formed by gravitational lensing near the event horizon. Recent imaging of the BH shadow in Sagittarius A* (Sgr A*) by the Event Horizon Telescope (EHT) \cite{shadow} has provided a groundbreaking test of strong-field gravity. Studies have examined BH shadows in numerous theoretical scenarios, including regular BHs, string-theory-inspired spacetimes, and alternative gravity models. Building upon these developments, the present work explores BH solutions embedded within DM halos, using the DM density profile $\rho_{\mathrm{DM}}$ from \cite{mod1}. By employing the spacetime construction techniques detailed in \cite{AA01, DM+materiaUsual, DM VC}, and incorporating shadow-related constraints from \cite{shadow}, we investigate how the presence of DM alters both shadow morphology and thermodynamic behavior of CBHs.

This work is organized as follows: In Section~\ref{sec2}, we present the generalized spacetime metric, the adopted DM density profile, and the Kretschmann scalar. Section~\ref{sec:solucoes_gerais} introduces two distinct BH solutions incorporating DM halos. Section~\ref{sec:propriedades} provides numerical estimates for the velocity, core radius, and BH mass, followed by an analysis of the properties of both solutions. Finally, Section~\ref{sec:concl} summarizes our conclusions.

All solutions considered below are static and spherically symmetric, with a metric signature $(-,+,+,+)$ and geometrized units ($G = c = 1$).

%%%%%%%%%%%%%%%%%%%%%%%%%%%%%%%%%%%%%%%%%%%%%%%%%%%%%%%%%%%%%%%%
%\section{Spacetime metric}\label{sec2}
%%%%%%%%%%%%%%%%%%%%%%%%%%%%%%%%%%%%%%%%%%%%%%%%%%%%%%%%%%%%%%%%

%%%%%%%%%%%%%%%%%%%%%%%%%%%%%%%%%%%%%%%%%%%%%%%%%%%%%%%%%%%%%%%%
\section{Spacetime Structure and Dark Matter Density Profile}\label{sec2}
%%%%%%%%%%%%%%%%%%%%%%%%%%%%%%%%%%%%%%%%%%%%%%%%%%%%%%%%%%%%%%%%

Consider the general metric expressed as
\begin{equation}
	ds^2 = -f(r)\,dt^2 + \frac{1}{g(r)}\,dr^2 + r^2\,d\Omega^2 \label{metrica} \ ,
\end{equation}
where the angular surface element in spherical coordinates is given by
$d\Omega^2 = d\theta^2 + \sin^2{\theta}\,d\phi^2$.

In the general case, $f(r)$ and $g(r)$ are distinct functions. However, throughout this work, we assume $f(r) = g(r)$ to simplify the calculations. We consider two approaches based on different methods: the first follows the framework proposed in \cite{AA01}, and the second is based on the treatment presented in \cite{DM+materiaUsual}.
In both methodologies, the resulting solutions reduce to an extension of the standard Schwarzschild black hole solution in the absence of a dark matter (DM) halo \cite{schw1999}. 

To describe the DM distribution, we adopt the density profile $\rho_{\mathrm{DM}}(r)$ introduced in \cite{mod1}, which is defined as
\begin{equation}
	\rho_{\mathrm{DM}}(r) = \frac{V_c^2}{4\pi G}\,\frac{3a^2 + r^2}{(a^2 + r^2)^2} \label{densidade} \ ,
\end{equation}
where $a$ denotes the characteristic core radius of the halo, and $V_c$ is the critical tangential velocity associated with the DM distribution.

Several properties of the BH solutions can be analyzed by the Kretschmann scalar $K = R_{\mu\nu\alpha\beta}R^{\mu\nu\alpha\beta}$, where $R_{\mu\nu\alpha\beta}$ denotes the Riemann tensor \cite{Kretsc.2021}. For the metric~(\ref{metrica}), with $f(r) = g(r)$, the scalar $K$ has the form
\begin{equation}
    K = f''(r)^2+\frac{4 f'(r)^2}{r^2}+\frac{4
   [f(r)-1]^2}{r^4}\,.
   \label{K_escalar_01}
\end{equation}

%%%%%%%%%%%%%%%%%%%%%%%%%%%%%%%%%%%%%%%%%%%%%%%%%%%%%%%%%%%%%%%%
\section{Black Hole Solutions with Dark Matter Halo}
\label{sec:solucoes_gerais}
%%%%%%%%%%%%%%%%%%%%%%%%%%%%%%%%%%%%%%%%%%%%%%%%%%%%%%%%%%%%%%%%

%%%%%%%%%%%%%%%%%%%%%%%%%%%%%%%%%%%%%%%%%%%%%%%%%%%%%%%%%%%%%%%%
\subsection{First Solution}\label{subs_model_01}
%%%%%%%%%%%%%%%%%%%%%%%%%%%%%%%%%%%%%%%%%%%%%%%%%%%%%%%%%%%%%%%%

To define the metric, we follow the procedure of \cite{AA01}, where the authors include the DM in the energy-momentum tensor $T_{\mu\nu}$. Thus, considering a Schwarzschild-type solution, we change the metric coefficients from Eq.~(\ref{metrica}): $f(r)\rightarrow F(r)=f(r)+F_{1}(r)$ and $g(r)\rightarrow G(r)=g(r)+F_{2}(r)$, which results in the following metric
\begin{equation}
    ds^2 = -\left[f(r)+F_1(r)\right]dt^2+\frac{1}{\left[g(r)+F_2(r)\right]}dr^2+r^2d\Omega^2 \label{metrica_mod}.
\end{equation}

The Einstein field equations \cite{Wald-RG1984} for this BH system are expressed as follows
\begin{equation}
    R_{\mu\nu}-\frac{1}{2}g_{\mu\nu}R=k^2\left(T_{\mu\nu}+T_{\mu\nu}^{(DM)}\right)\label{eq.einstein_mod},
\end{equation}
where $T_{\mu\nu}^{(DM)}$ is the contribution of DM, and $R_{\mu\nu}$ is the Ricci tensor, $R$ is the scalar curvature and $g_{\mu\nu}$ is the metric tensor. 

Now, inserting the space-time metric (\ref{metrica_mod}) into the Einstein field equation (\ref{eq.einstein_mod}), yields \cite{AA01}
{\small\begin{equation}
\left[g(r) + F_2(r)\right] \left[ \frac{1}{r^2} + \frac{g'(r) + F_2'(r)}{r \left[g(r) + F_2(r)\right]} \right] = g(r) \left[\frac{1}{r^2} + \frac{g'(r)}{r g(r)} \right],
\label{eqD1}
\end{equation}}
{\small\begin{equation}
\left[g(r) + F_2(r)\right] \left[ \frac{1}{r^2} + \frac{f'(r) + F_1'(r)}{r \left[f(r) + F_1(r)\right]} \right] = g(r) \left[\frac{1}{r^2} + \frac{f'(r)}{r f(r)} \right],\label{eqD2}
\end{equation}}
respectively.
By solving Eq.~(\ref{eqD1}) and Eq.~(\ref{eqD2}) we obtain
\begin{equation}
    F(r)=-\exp\left\{\scalebox{1.7}{$\int$} \left[ \frac{g(r)}{ g(r) - \frac{2M}{r}} \left( \frac{1}{r} + \frac{f'(r)}{f(r)} \right) - \frac{1}{r} \right] \, dr\right\},
\end{equation}
and
\begin{equation}
    G(r)=g(r)-\frac{2M}{r} \ ,
\end{equation}
which leads to the following reformulation of the metric~(\ref{metrica_mod}),
{\small\begin{eqnarray}
       ds^2 &=& -\exp\left\{\scalebox{1.7}{$\int$} \left[ \frac{g(r)}{ g(r) - \frac{2M}{r}} \left( \frac{1}{r} + \frac{f'(r)}{f(r)} \right) - \frac{1}{r} \right] \, dr\right\} dt^2 
       \nonumber \\
&&  \hspace{1cm}
       + \left[ g(r) - \frac{2M}{r} \right]^{-1} \, dr^2 + r^2 d\Omega^2 \label{Met_halo} .
\end{eqnarray}}
Equation~(\ref{Met_halo}) describes a more general Schwarzschild-type solution for BHs with a DM halo. To return to the scenario without DM, we simply set $f(r)=g(r)=1$ as shown in \cite{AA01}.

We now consider $f(r)=g(r)$ and use the mass relation $m(r)$ and the density $\rho(r)$ with the tangential velocity $v_{tg}(r)$ and the function $f(r)$, both of which are represented in \cite{baryonic}.
The first relationship is
\begin{equation}
    m(r)=4\pi \int^{r}_{0}\rho(h)h^2dh \ ,
\end{equation}
in which we use the density from Eq.~(\ref{densidade}), which leads to
\begin{equation}
    m(r)=\frac{V_c^2r^3}{(a^2+r^2)}\label{massa} \ .
\end{equation}
The second relation is given by
\begin{equation}
    v_{tg}^2=\frac{m(r)}{r}\label{vtg01}
\end{equation}
and
\begin{equation}
v_{tg}^2(r)=\frac{d\ln{\sqrt{f(r)}}}{d\ln{r}}=r\frac{d\ln{\sqrt{f(r)}}}{dr}\label{vtg02} \ .
\end{equation}

Thus, using Eqs.~(\ref{massa})--(\ref{vtg02}), we deduce the function $f(r)$ as
\begin{equation}
f(r)=\gamma\left(a^2+r^2\right)^{V^2_c},   
\end{equation}
therefore,
\begin{equation}
F(r)=G(r)=\gamma\left(a^2+r^2\right)^{V^2_c} - \frac{2M}{r}\label{FG_primeiraSol},
\end{equation}
where $\gamma$ is an integration constant. Finally, the metric~(\ref{Met_halo}) becomes
\begin{eqnarray}
       ds^2 &=& -\left[ \gamma\left(a^2+r^2\right)^{V^2_c} - \frac{2M}{r} \right] dt^2 
       \nonumber \\
&&
+ \left[ \gamma\left(a^2+r^2\right)^{V^2_c} - \frac{2M}{r} \right]^{-1} \, dr^2 + r^2 d\Omega^2  \ ,\label{MetHalo_sol01}
\end{eqnarray}
which reduces to the Schwarzschild solution if $V_c=0$ and $\gamma=1$. In the next section we will discuss the conditions for the behavior of $\gamma$.

%%%%%%%%%%%%%%%%%%%%%%%%%%%%%%%%%%%%%%%%%%%%%%%%%%%%%%%%%%%%%%%%
\subsection{Second Solution}
%%%%%%%%%%%%%%%%%%%%%%%%%%%%%%%%%%%%%%%%%%%%%%%%%%%%%%%%%%%%%%%%

The second method used in this paper to define the metric functions follows the approach proposed by the authors of \cite{DM+materiaUsual}. We start from the expression
\begin{equation}
    F(r)=\frac{C}{r}+\sum_{i}f_i(r)\label{fi} ,
\end{equation}
where $C$ is a constant yet to be determined, and from the expression (which takes into account up to the $i$-th contribution to the metric function)
\begin{equation}
\frac{f_i(r)+rf'_{i}(r)}{r^2}=    8\pi \sum_{i} \rho^{(i)}(r)\,,
\label{densidade_fi}
\end{equation}
with reference to the following metric:
\begin{equation}
ds^{2} = -\left[1+F(r)\right] dt^{2} + \frac{1}{1+F(r)} dr^{2} + r^{2} d\Omega^{2}\label{metrica_met02}.
\end{equation}
Substituting the density~(\ref{densidade}) into Eq.~(\ref{densidade_fi}), we get
\begin{equation}
    \sum_{i}f_{i}=\frac{C_1}{r}+\frac{2V^2_cr^2}{(a^2+r^2)} ,
\end{equation}
where $C_1$ is an integration constant. Thus, from Eq.~(\ref{fi})
\begin{equation}
    F(r)=\frac{C_2}{r}+\frac{2V^2_cr^2}{c^4(a^2+r^2)}\label{f(r)_model02},
\end{equation}
with the constant $C_2 = C + C_1$. To construct an explicit equivalence to the Schwarzschild solution, we define $C_2 = -2M$, which leads to the following metric given by Eq.~(\ref{metrica_met02})
\begin{align}
ds^{2} = & -\left[1-\frac{2M}{r}+\frac{2V^2_cr^2}{(a^2+r^2)}\right] dt^{2} \nonumber \\
& + \left[1-\frac{2M}{r}+\frac{2V^2_cr^2}{(a^2+r^2)}\right]^{-1} dr^{2} + r^{2} d\Omega^2\label{metrica_completa_model02}.
\end{align}

%%%%%%%%%%%%%%%%%%%%%%%%%%%%%%%%%%%%%%%%%%%%%%%%%%%%%%%%%%%%%%%%
\section{Properties}\label{sec:propriedades}
%%%%%%%%%%%%%%%%%%%%%%%%%%%%%%%%%%%%%%%%%%%%%%%%%%%%%%%%%%%%%%%%

In this section we use the values for the mass $M$, the critical velocity $V_c$ and the critical radius $a$ as defined in \cite{mod1}, for the SMBH and DM Halo of NGC 4649 (M60). This defines the intervals described in Tab.~\ref{tab:parametros}, which include both International System (IS) units and geometrized units.
\par
In these intervals, we highlight two main groups of values: the first, which we refer to as Data I, comes from X-ray observations \cite{mod1}; and the second, which we refer to as Data II, refers to the mass dynamics profile obtained by the authors in \cite{mod1}. The values are presented as follows: $V_{cI}=13.68\times10^{-4}$, $a_{I}=30.86\times10^{19}\textnormal{m}$, and $M_{I}=5.17\times10^{12}\textnormal{m}$ for Data I, and $V_{cII}=18.35\times10^{-4}$, $a_{II}=46.29\times10^{19}\textnormal{m}$, and $M_{II}=6.65\times10^{12}\textnormal{m}$ for Data II.
\begin{table}[h]
    \centering
\renewcommand{\arraystretch}{1.2}
    \begin{tabular}{lcc}
        \toprule
        \textbf{Parameter} & \textbf{Minimum} & \textbf{Maximum} \\
        \midrule
        $V_c$[km/s] (IS) & $300$ & $1100$ \\
        $V_c$ (Geom.) & $10 \times 10^{-4}$ & $36.7 \times 10^{-4}$ \\
        $a$[kpc] (IS) & $5$ & $90$ \\
        $a$[m] (Geom.) & $15.43 \times 10^{19}$ & $277.74 \times 10^{19}$ \\
        $M$[$M_{\odot}$] (IS) & $2.5 \times 10^{9}$ & $6 \times 10^{9} $ \\
        $M$[m] (Geom.) & $3.69 \times 10^{12}$ & $8.86 \times 10^{12}$ \\
        \bottomrule
    \end{tabular}
    \caption{Parameters $V_c$, $a$ and $M$ in different units with minimum and maximum values.}
    \label{tab:parametros}
\end{table}
\par
According to \cite{mod1}, the mass in the absence of a halo assumes the value $M_0=4.3\times10^{9}M_{\odot}$, which corresponds to $M_0=6.35\times10^{12}\textnormal{m}$.
\par
Another measure we will use to check the properties of the solutions is the distance $r_O$ from an observer on Earth to M60, which is given by $r_O=15.7\times10^{3}$ kpc, or equivalently $r_O=48.45\times10^{22}$ m.

%%%%%%%%%%%%%%%%%%%%%%%%%%%%%%%%%%%%%%%%%%%%%%%%%%%%%%%%%%%%%%%%
\subsection{First Solution}\label{subsec:gtt_01}
%%%%%%%%%%%%%%%%%%%%%%%%%%%%%%%%%%%%%%%%%%%%%%%%%%%%%%%%%%%%%%%%

%%%%%%%%%%%%%%%%%%%%%%%%%%%%%%%%%%%%%%%%%%%%%%%%%%%%%%%%%%%%%%%%
\subsubsection{Event Horizon}
%%%%%%%%%%%%%%%%%%%%%%%%%%%%%%%%%%%%%%%%%%%%%%%%%%%%%%%%%%%%%%%%

The condition for determining the event horizon $r_{h}$ is the solution of the equation $F(r_h) = 0$. With the first solution (\ref{FG_primeiraSol})
\begin{equation}
F(r)=\gamma\left(a^2+r^2\right)^{V^2_c} - \frac{2M}{r} \ \label{F_01}.
\end{equation}
we have considerable difficulty in obtaining an analytical solution.
However, we may represent the behavior of $F(r)$ with respect to $r$ and calculate $r_{h}$ using defined values for the parameters $M$, $V_c$ and $a$.
\par 
The constant $\gamma$ is initially not defined. However, to restore the Schwarzschild solution, it is necessary that $\gamma\left(a^2 + r^2\right)^{V_c^2} \to 1$. So we consider small values of $r$ so that $r^2/a^2 \ll 1$, which leads to the following result
\begin{equation}
\gamma\rightarrow\gamma_c = a^{-2V_c^{2}}\label{gamma_critico}\ .
\end{equation}
The constant $\gamma_c$ adjusts the halo metric for small $r$ so that it returns to a solution that behaves similarly to the Schwarzschild solution without a halo. Equation~(\ref{gamma_critico}) shows that the constant $\gamma_c$ in this region depends directly on $a$ and $V_c$ and provides a smooth transition between the Schwarzschild spacetime and the halo metric. Due to the size of $a$ and $V_c$, their influence is not significant in the region near the BH. $\gamma_c$ becomes more relevant for higher values of $V_c$ and $a$, where $\gamma_c < 1$.

%  scale=0.28
\begin{figure}[htb!]
\includegraphics[width=\linewidth]{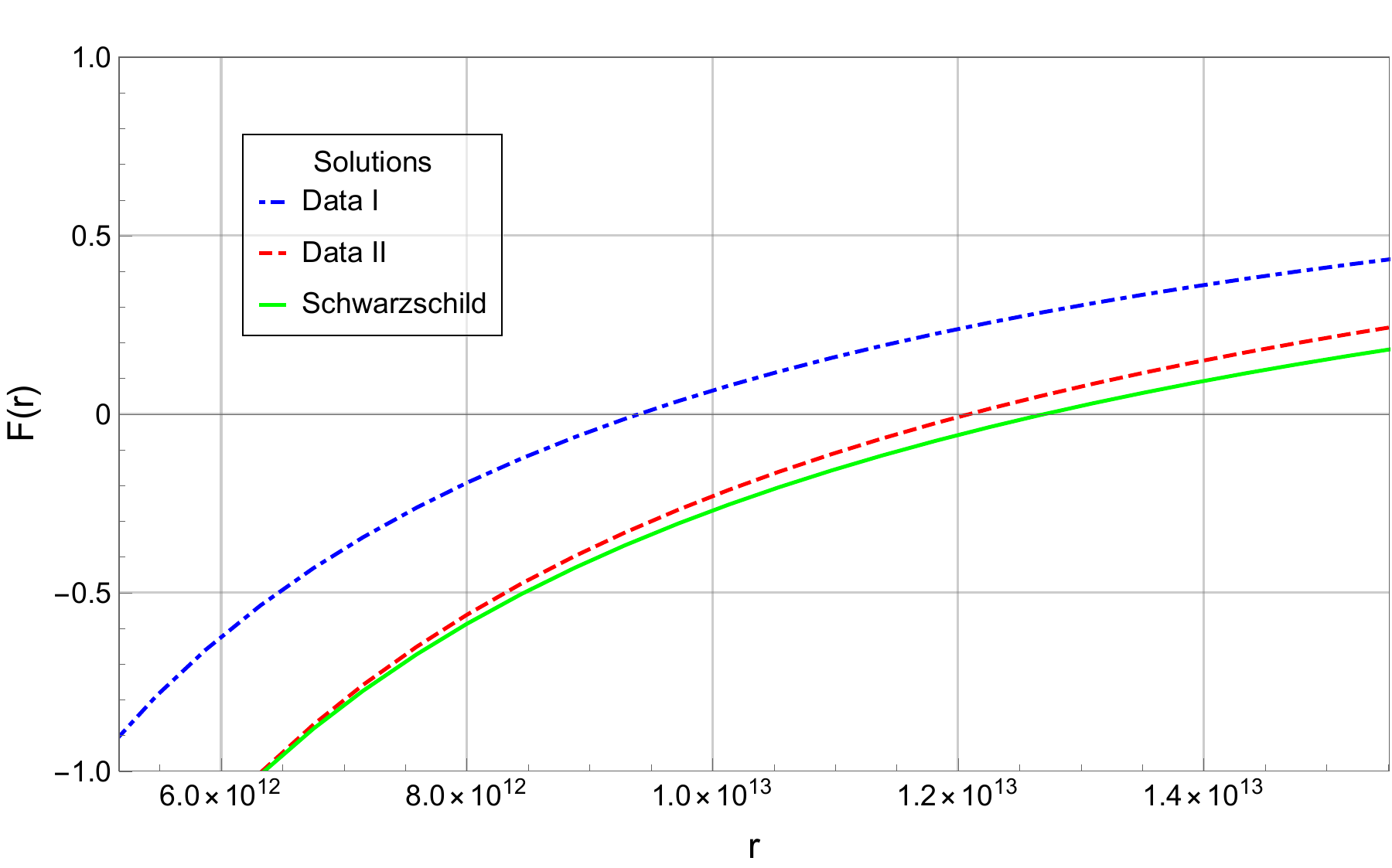}
\caption{Graphical representation of $F(r)$, Eq.~(\ref{F_01}), for the parameter sets: ($V_{cI}, a_I, M_I, \gamma_0$) given as Data I; ($V_{cII}, a_{II}, M_{II}, \gamma_0$) given as Data II; and the solution of the Schwarzschild-type without halo $(0, 0, M_0, 1)$.
} 
\label{fig:F(r)_vs_r_model01}
\end{figure}

For the following calculations, we define $\gamma_0 = 1.1$ and analyze the behavior of $F(r)$, as shown in Fig.~\ref{fig:F(r)_vs_r_model01}, which indicates the existence of an event horizon $r_h$. The values obtained for $r_h$ are: $r_{hI} = 9.398\times10^{12}\,\textnormal{m}$, $r_{hII} = 1.209\times10^{13}\,\textnormal{m}$, and $r_{h}^{\textnormal{(sch)}} = 1.27\times10^{13}\,\textnormal{m}$, respectively for the data sets Data I, Data II and for solutions of Schwarzschild-type, i.e. without a halo.

Figure~\ref{fig:F(r)_vs_r_model01} shows that the presence of the DM halo exerts a considerable influence on the event horizon radius and reduces it. This effect implies that a larger mass is required for the radius to reach the same size as $r_h^{\textnormal{(sch)}}$, which corresponds to the solution of the Schwarzschild-type.

%%%%%%%%%%%%%%%%%%%%%%%%%%%%%%%%%%%%%%%%%%%%%%%%%%%%%%%%%%%%%%%%
\subsubsection{Kretschmann Scalar}
%%%%%%%%%%%%%%%%%%%%%%%%%%%%%%%%%%%%%%%%%%%%%%%%%%%%%%%%%%%%%%%%

\begin{figure}[htb!]
\includegraphics[width=\linewidth]{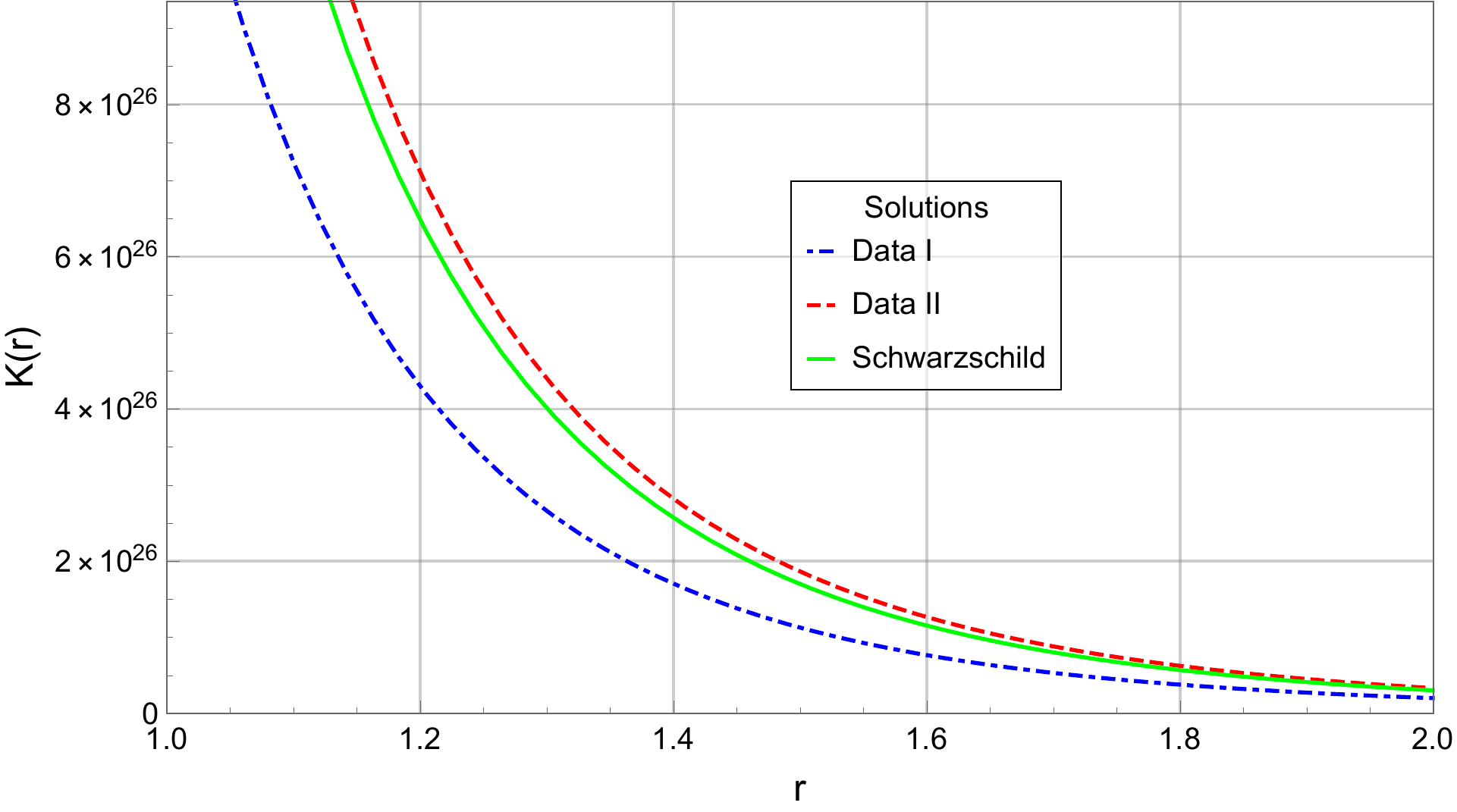}
\caption{Graphical representation of the scalar $K(r)$, from Eq.~(\ref{K_model_01_explicito}), for the values: $(V_{cI}, a_I, M_I, \gamma_0)$ given by Data I, $(V_{cII}, a_{II}, M_{II}, \gamma_0)$ given by Data II, and the solution of Schwarzschild-type without halo $(0,0, M_0, 1)$.} \label{fig:K(r)_vs_r_model01}
\end{figure}

Assuming that $f(r) = g(r)$, so that $F(r) = G(r)$, the Kretschmann scalar $K$ is obtained from Eq.~(\ref{K_escalar_01})
\begin{equation}
    K = F''(r)^2+\frac{4 F'(r)^2}{r^2}+\frac{4
   [F(r)-1]^2}{r^4} \,.
\end{equation}
Substitute Eq.~(\ref{FG_primeiraSol}), yeilds the following expression
\begin{eqnarray}
	K &=& \frac{4}{r^2} \left[2 \gamma r V_{c}^2\left(a^2+r^2\right)^{V_{c}^2-1}
		+\frac{2M}{r^2}\right]^2
		\nonumber \\
		&&
		+\frac{4}{r^4}\left[\gamma
		\left(a^2+r^2\right)^{V_{c}^2}-\frac{2M}{r}-1\right]^2
		\nonumber \\
		&&
		+ \Big[4 \gamma r^2 V_{c}^2
	\left(V_c^2-1\right)
	\left(a^2+r^2\right)^{V_{c}^2-2}
	\nonumber
	\\
	&& +2 \gamma V_c^2
	\left(a^2+r^2\right)^{V_{c}^2-1}-\frac{4M}{
		r^3}\Bigg]^2 \,.
		\label{K_model_01_explicito}
\end{eqnarray}  
The asymptotic behavior of the curvature, which is represented by the scalar $K$ from Eq.~(\ref{K_model_01_explicito}), can be analyzed using the limit values:
\begin{equation}
    \lim_{r \to \infty} K\rightarrow 0
\end{equation}
and
\begin{equation}
    \lim_{r \to 0} K\rightarrow \infty ,
\end{equation}
which indicates a flat spacetime and a singularity. This is depicted in Fig.~\ref{fig:K(r)_vs_r_model01}, where the scalar $K$ shows a decreasing behavior with increasing radius, with a tendency towards a flat spacetime.

We emphasize the sensitivity of the scalar $K$ to the parameters $V_c$ and $a$. Its curve is slightly shifted, indicating a change in the intensity of the curvature in the presence of the halo.

%%%%%%%%%%%%%%%%%%%%%%%%%%%%%%%%%%%%%%%%%%%%%%%%%%%%%%%%%%%%%%%%
\subsubsection{Black Hole Shadow Radius}
%%%%%%%%%%%%%%%%%%%%%%%%%%%%%%%%%%%%%%%%%%%%%%%%%%%%%%%%%%%%%%%%

The existence of a region outside the BH containing a photon sphere allows an observer at distance $r_O$ to detect the shadow of this BH \cite{shadow}. Using the metric coefficient from Eq.~(\ref{metrica}), and following the approaches outlined in \cite{shadow,Perlick:2021aok}, we define a function $h(r)$ as
\begin{equation}
    h(r)=\sqrt{\frac{r^2}{f(r)}} .
\end{equation}
The radius of the photon sphere $r_{ph}$ can therefore be determined by
\begin{equation}
    \left[\frac{dh^2(r)}{dr}\right]_{r_{ph}} = 0 ,
\end{equation}
this implies that
\begin{equation}
    \left[F(r)-\frac{r}{2}\frac{dF(r)}{dr}\right]_{r_{ph}}=0\label{r_ph__metodo}.
\end{equation}
The shadow radius $r_{sh}$ can be determined by the expression:
\begin{equation}
    r_{sh}=r_{ph}\sqrt{\frac{F(r_{O})}{F(r_{ph})}}\label{r_sh_GERAL_naoplana} \ ,
\end{equation}
where $r_{O}$ is the distance between a static observer and the BH. For the asymptotically flat solution, where $F(r_O)=1$, we have
\begin{equation}
    r_{sh}=\frac{r_{ph}}{\sqrt{F(r_{ph})}}\label{r_sh_GERAL} \ .
\end{equation}
Just like the radius of the photon sphere, the radius of the shadow must also be larger than the radius of the event horizon \cite{shadow}.
\par
To obtain $r_{ph}$ in the first solution, we substitute Eq.~(\ref{FG_primeiraSol}) into Eq.~(\ref{r_ph__metodo}), and then solve the following equation:
\begin{equation}
(a^2+r^2)^{V_c^2}-r^2V_c^2(a^2+r^2)^{V_c^2-1}=\frac{2M}{\gamma r}.\label{r_phSOL1}
\end{equation}
To reduce the algebraic complexity, we rewrite Eq.~(\ref{r_phSOL1}) under the approximations $V_c \ll 1$ and $r \ll a$, which are consistent with the numerical results for Data Sets I and II. In these cases, $a \gg 1$ and $V_c \ll 1$. These approximations are further justified by: (i) the inaccuracy of the parameters at large distances, as noted in \cite{mod1}, and (ii) the dominance of stellar mass over dark matter at the halo's effective radius, as considered by the cited authors.
In view of this, from Eq.~(\ref{r_phSOL1}), we deduce:
\begin{equation}
r_{ph}=\frac{2M}{\gamma }a^{-2V_c^2}\label{r_ph_SOL1_r_pequeno}.
\end{equation}
This approximation is consistent with the values of the parameters $V_c$ and $a$. We calculate the radius of the photon sphere, $r_{ph,n}$, numerically without algebraic approximation from Eq.~(\ref{r_phSOL1}) and compare it with Eq.~(\ref{r_ph_SOL1_r_pequeno}). The resulting deviations are $1 - \frac{r_{ph}}{r_{ph,n}} = 1.92 \times 10^{-16}$ for Data I and $\frac{r_{ph}}{r_{ph,n}} - 1 = 0.17 \times 10^{-16}$ for Data II. This supports the validity of Eq.~(\ref{r_ph_SOL1_r_pequeno}).

According to \cite{shadow}, the radius of the event horizon must be smaller than the radius of the photon sphere, $r_{h} < r_{ph}$. Thus, by analysing Eq.~ (\ref{FG_primeiraSol}) in the regime $r \ll a$, we get
\begin{equation}
 r_{h}=\frac{2M}{\gamma }a^{-2V_c^2} ,\label{rh_model01_}
\end{equation}
which shows that this criterion is met by the model. If $\gamma = \gamma_c$, as given in Eq.~(\ref{gamma_critico}), we obtain the radii of the photon sphere and the event horizon corresponding to the Schwarzschild solution, $r_{ph} = 3M$ and $r_{h} = 2M$. The same condition for $\gamma$ also applies to the metric given by Eq.~(\ref{MetHalo_sol01}).

To obtain the shadow radius, we use Eq.~(\ref{r_ph_SOL1_r_pequeno}) in Eq.~(\ref{r_sh_GERAL_naoplana}), which leads to the following result
\begin{equation}
r_{sh}=\frac{3a^{-2V_c}M[F(r_O)]^{1/2}}{\gamma^2\left[ \left(a^2+\frac{9M^2}{\gamma^2}a^{-4V_c}\right)^{V_c^2}-\frac{2}{3}a^{2V_c}\right]^{1/2}}\label{sombra_r_pequeno}.
\end{equation}
To obtain the shadow radius of the Schwarzschild solution, $r_{sh} = 3\sqrt{3}M$, from Eq.~(\ref{sombra_r_pequeno}), it is sufficient to set $V_c = 0$, and consequentely, $\gamma = 1$.

The percentage difference between the observed shadow radius $r_{sh,ob}$ and the expected shadow radius for the Schwarzschild solution is quantified by
\begin{equation}
    \delta = \frac{r_{sh,ob}}{3\sqrt{3}M}-1 ,
\end{equation}
given the data obtained by two teams, Keck and VLTI \cite{shadow}. Since the groups are uncorrelated, the mean of the estimates of $\delta$ yields the following values:
\begin{equation}
\begin{aligned}
-0.125 \lesssim \delta \lesssim 0.005 \\
-0.19 \lesssim \delta \lesssim 0.07 \ ,
\end{aligned}
\end{equation}
which correspond to $1\sigma$ and $2\sigma$ respectively. This results in the following restrictions for the shadow radius, expressed in its ratio to the mass,
\begin{equation}
    4.55\lesssim r_{sh}/M\lesssim5.22
\end{equation}
for $1\sigma$, and
\begin{equation}
    4.21\lesssim r_{sh}/M\lesssim5.56
\end{equation}
for $2\sigma$. We will use these intervals to assess the validity of certain aspects of the metric models.

\begin{figure}[htb!]
\includegraphics[width=\linewidth]{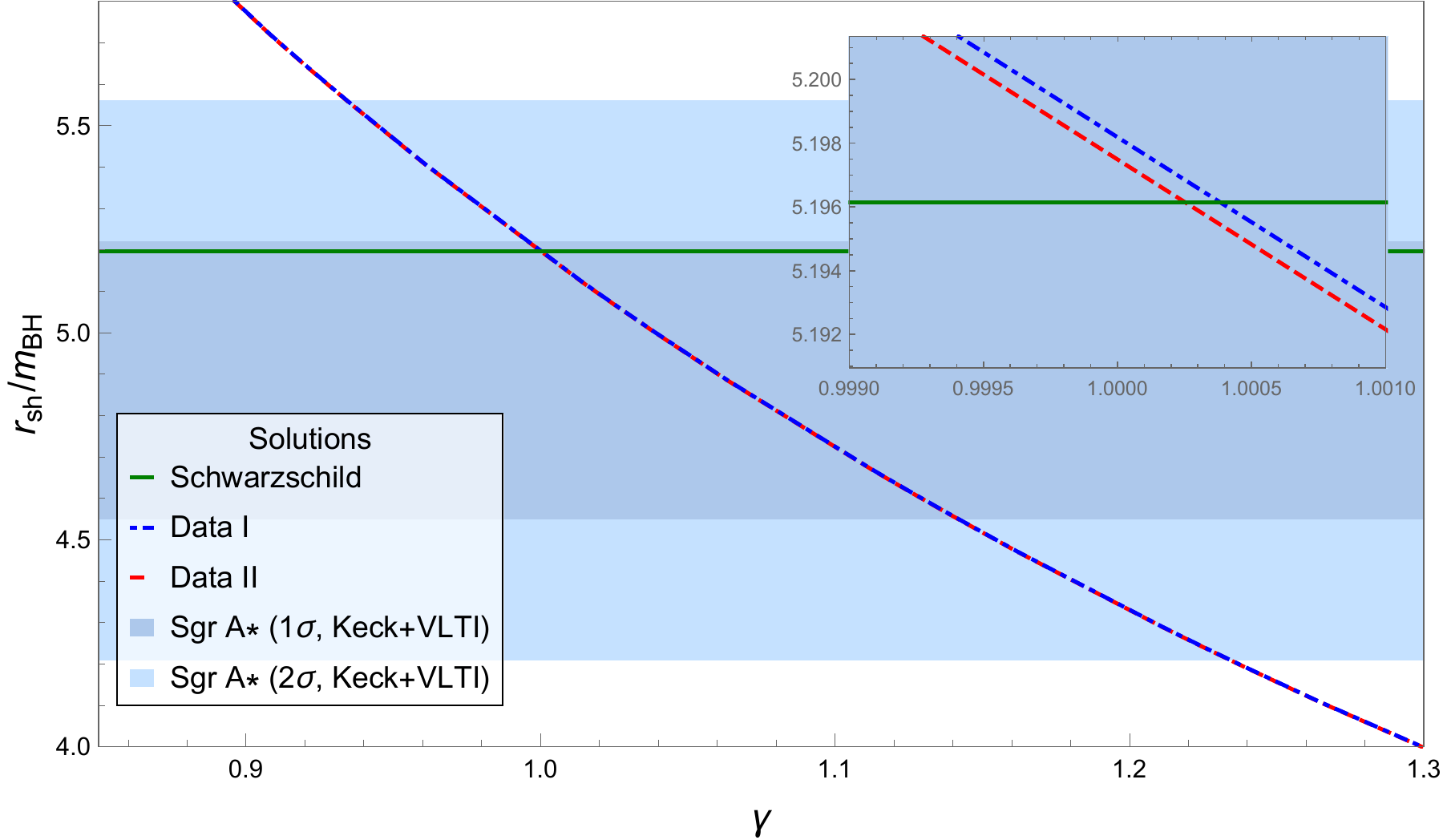}
\caption{Graphical representation of the ratio $r_{sh}/M$ from Eq.~(\ref{sombra_r_pequeno}) for values of $\gamma$ within the measurement intervals corresponding to $1\sigma$ and $2\sigma$. For the parameter sets: ($V_{cI},a_I,M_I,\gamma$), given by Data I, ($V_{cII},a_{II},M_{II},\gamma$), given by Data II, and the solution of type SS without halo $(0,0,M_0,1)$.}
\label{fig:rshadow_vs_gamma_model01}
\end{figure}

The behavior of the shadow radius, as described in Eq.~(\ref{sombra_r_pequeno}), is analyzed as a function of the values of the constant $\gamma$, as shown in Fig.~\ref{fig:rshadow_vs_gamma_model01}. For both pairs of values, there is an interval of $\gamma$ within the valid range (blue areas) that includes $\gamma_0$. This shows that $\gamma$ can be treated as a new parameter in addition to $V_c$ and $a$. The $\gamma$ intervals obtained for $r_{sh}/M$ are $0.934408\leq\gamma_{I}\leq1.23404$ and $0.934282\leq\gamma_{II}\leq1.23387$ for Data I and II, respectively. In fact, the model agrees with the $1\sigma$ and $2\sigma$ intervals, which supports its validity and the approximations made for $r_{ph}$.

The subfigure of Fig.~\ref{fig:rshadow_vs_gamma_model01} shows a slight variation of the $\gamma$ interval in response to changes in the other parameters. The increase or decrease of $r_{sh}$ corresponds to the selected value of $\gamma$. For the originally chosen value, $\gamma_0$, we have $r_{sh}|_{\gamma_0}<r_{sh}|_{\gamma=1}$, which is consistent with the decrease in $r_h$ described in Fig.~\ref{fig:F(r)_vs_r_model01}, since $r_h|_{\gamma_0}<r_{sh}|_{\gamma_0}$.

With good approximation, we can obtain an expression for $\gamma$ if we consider the scenario where the observer is sufficiently far away from the BH, i.e. $F(r_O)\to1$. This behavior is satisfied for the Schwarzschild metric when $r_O\gg M$. If we assume this relationship and $r_O\gg a$, we obtain from Eq.~(\ref{FG_primeiraSol}) that
\begin{equation}
\gamma\to\gamma_O\approx r_O^{-2V^2_c} \label{gamma_aproximado}\ ,
\end{equation}
in which, if $V_c=0 \Rightarrow \gamma_O=1$, the Schwarzschild metric is restored. Thus, $\gamma_O$ can be defined as a function of the halo velocity and the distance of the observer. For Eq.~(\ref{gamma_aproximado}) we have: $\gamma_{OI}\approx0.999796$ and $\gamma_{OII}\approx0.999633$. These values agree well with the definition of $\gamma_c$ in Eq.~(\ref{gamma_critico}), where we get for this expression: $\gamma_{cI}\approx0.999823$ and $\gamma_{cII}\approx0.99968$.

Since we have previously shown that the size of $r_{sh}$ is sensitive to $\gamma$, the shadow radius for a distant observer, $r\gg a$, becomes visually larger compared to the shadow radius perceived by an observer in regions close to the BH, $r\ll a$, because $\gamma_O\lesssim\gamma_c$.

%%%%%%%%%%%%%%%%%%%%%%%%%%%%%%%%%%%%%%%%%%%%%%%%%%%%%%%%%%%%%%%%
\subsubsection{Thermodynamics}
%%%%%%%%%%%%%%%%%%%%%%%%%%%%%%%%%%%%%%%%%%%%%%%%%%%%%%%%%%%%%%%%

The entropy $S$ of a BH is introduced by its ratio to the area of the event horizon $A$ \cite{S-Hawking1975},
\begin{equation}
    S=\frac{A}{4} ,
\end{equation}
where $A = 4\pi r_{h}^2$ \cite{Wald-RG1984, Bardeen1973}, so that
\begin{equation}
    S=\pi r^2_{h}\label{S}.
\end{equation}
We obtain the relationship between entropy and mass by setting $F(r_{h})=0$ in Eq.~(\ref{FG_primeiraSol}) and using Eq.~(\ref{S}), which yields
\begin{equation}
M(S,\gamma,a)=\frac{\gamma \sqrt{S}}{2\sqrt{\pi}}\left(a^2+\frac{S}{\pi}\right)^{V_c^2}\label{Massa_sol01}    \ .
\end{equation}

\begin{figure}[htb!]
\includegraphics[width=\linewidth]{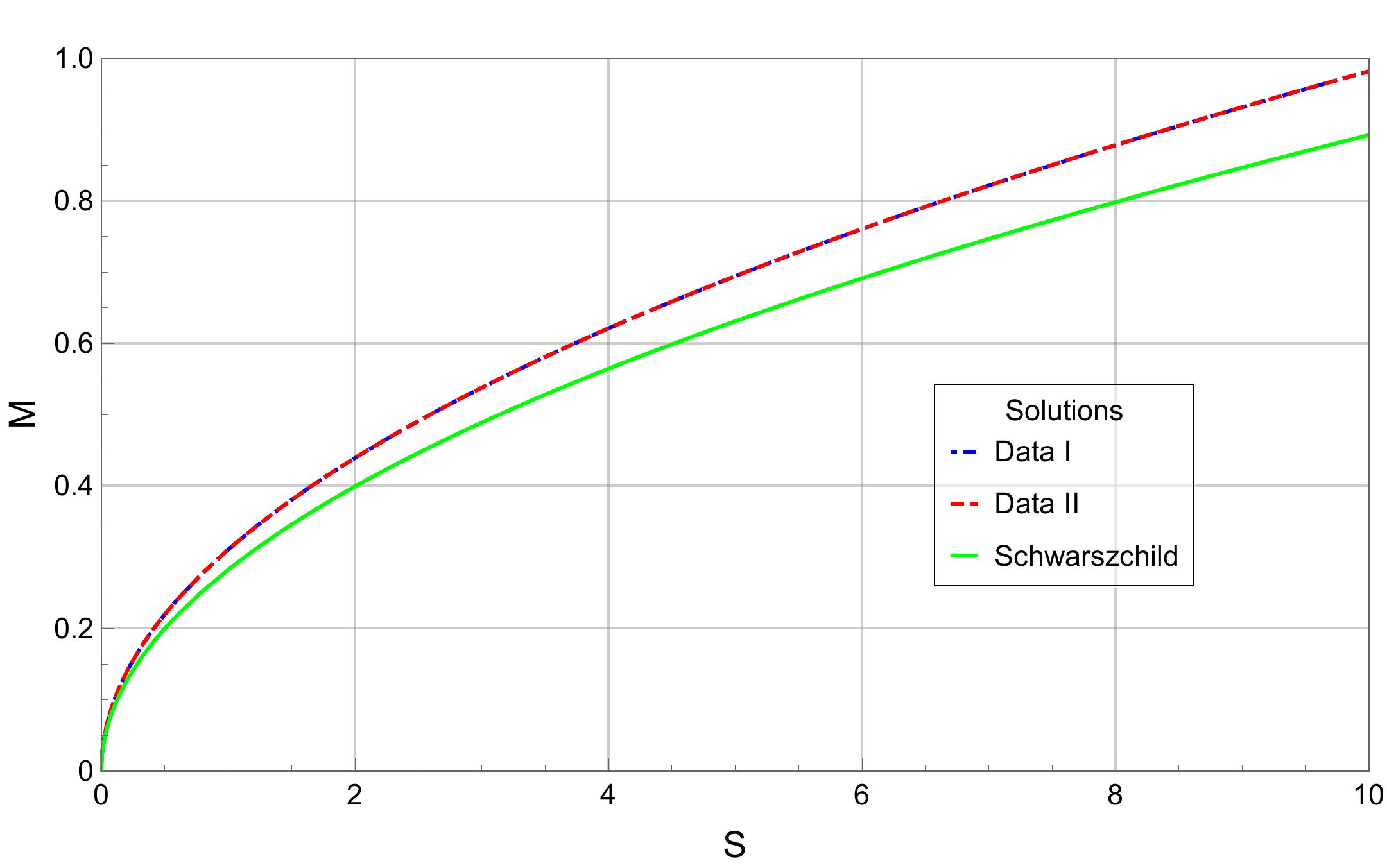}
\caption{Graphical representation of the mass $M(S)$, from Eq.~(\ref{Massa_sol01}). For the values: ($V_{cI},a_I,\gamma_0$), given by Data I, and ($V_{cII},a_{II},\gamma_0$), given by Data II, and the Schwarzschild solution without halo (0,0,1).
} 
\label{fig:Massa_vs_S_model01}
\end{figure}

Figure~\ref{fig:Massa_vs_S_model01} shows the behavior of mass in relation to entropy. Although in general $M^2 \propto S$ is conserved due to the low velocity, the intensity of the entropy decreases with the presence of the halo at the same mass value. This reflects the response to the reduction of $r_h$ since $S \propto r_h^2$. However, the mass is changed by the presence of the halo \cite{mod1}.
These relationships indicate that the BH with halo reacts with a lower entropy despite its greater mass, suggesting an increase in its mass density.
% scale=0.23

\begin{figure}[htb!]
\includegraphics[width=\linewidth]{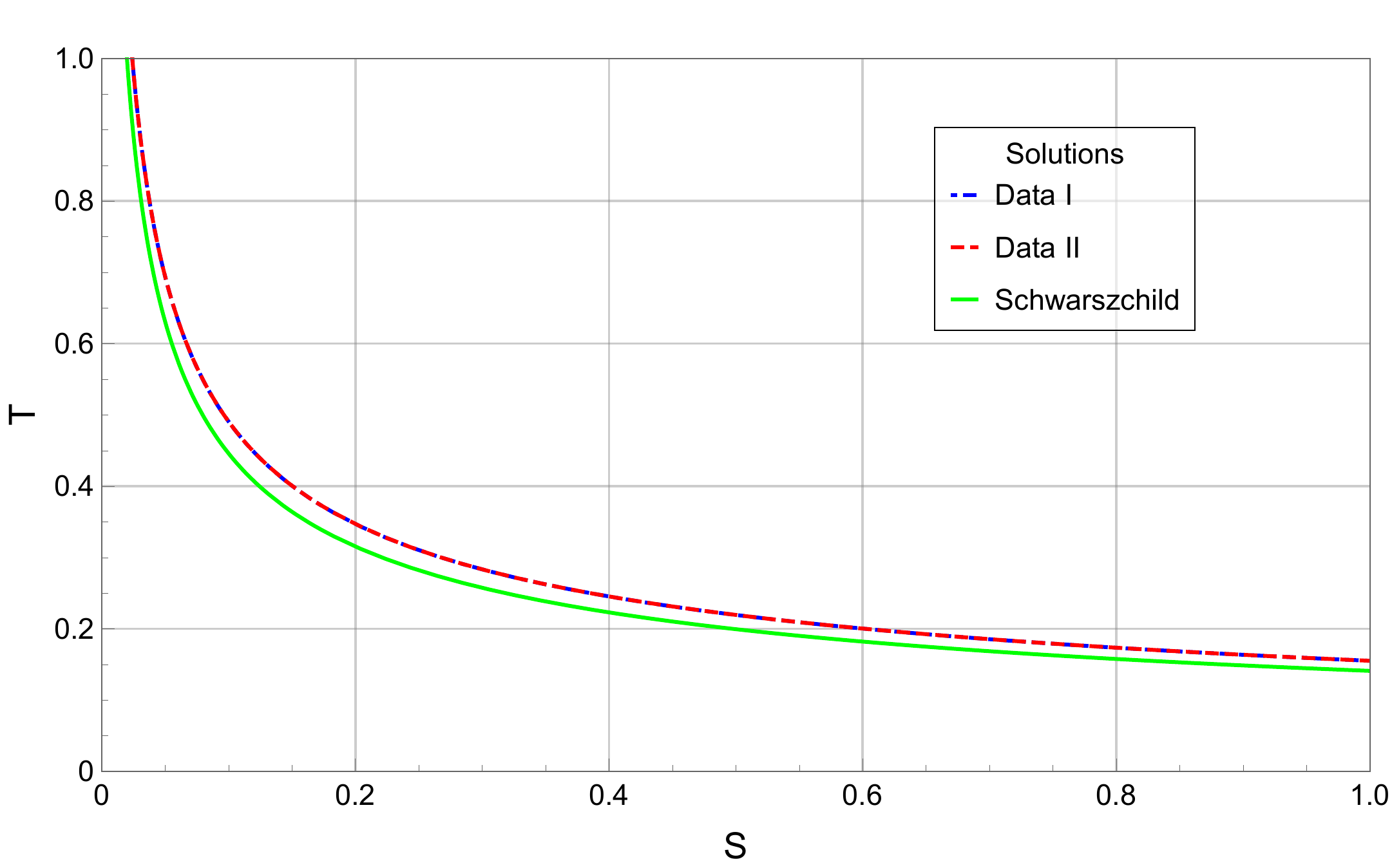}
\caption{Graphical representation of the temperature $T(S)$ from Eq.~(\ref{temperatura_sol01}). For the values: ($V_{cI},a_I,\gamma_0$), given by Data I, and ($V_{cII},a_{II},\gamma_0$), given by Data II, and the solution of type SS without halo (0,0,1).} 
\label{fig:T_vs_S_model01}
\end{figure}

To obtain the temperature $T$, we use $T = \frac{\partial M}{\partial S}$, which leads to the following result
\begin{equation}
    T(S,\gamma,a) = \frac{\gamma \pi
   ^{-V_c^2-\frac{1}{2}}
   \left(\pi 
   a^2+S\right)^{V_c^2-1}
   \left(\pi  a^2+2 S
   V_c^2+S\right)}{4 \sqrt{S}}\label{temperatura_sol01} .
\end{equation}
Figure~\ref{fig:T_vs_S_model01}, from Eq.~(\ref{temperatura_sol01}), shows that a BH with a halo has a slight increase in temperature compared to the case without a halo. The general behavior $T^{-2} \propto S$ is maintained in both scenarios, indicating that the influence of the halo is of the intensity of the temperature drop.

\begin{figure}[htb!]
\includegraphics[width=\linewidth]{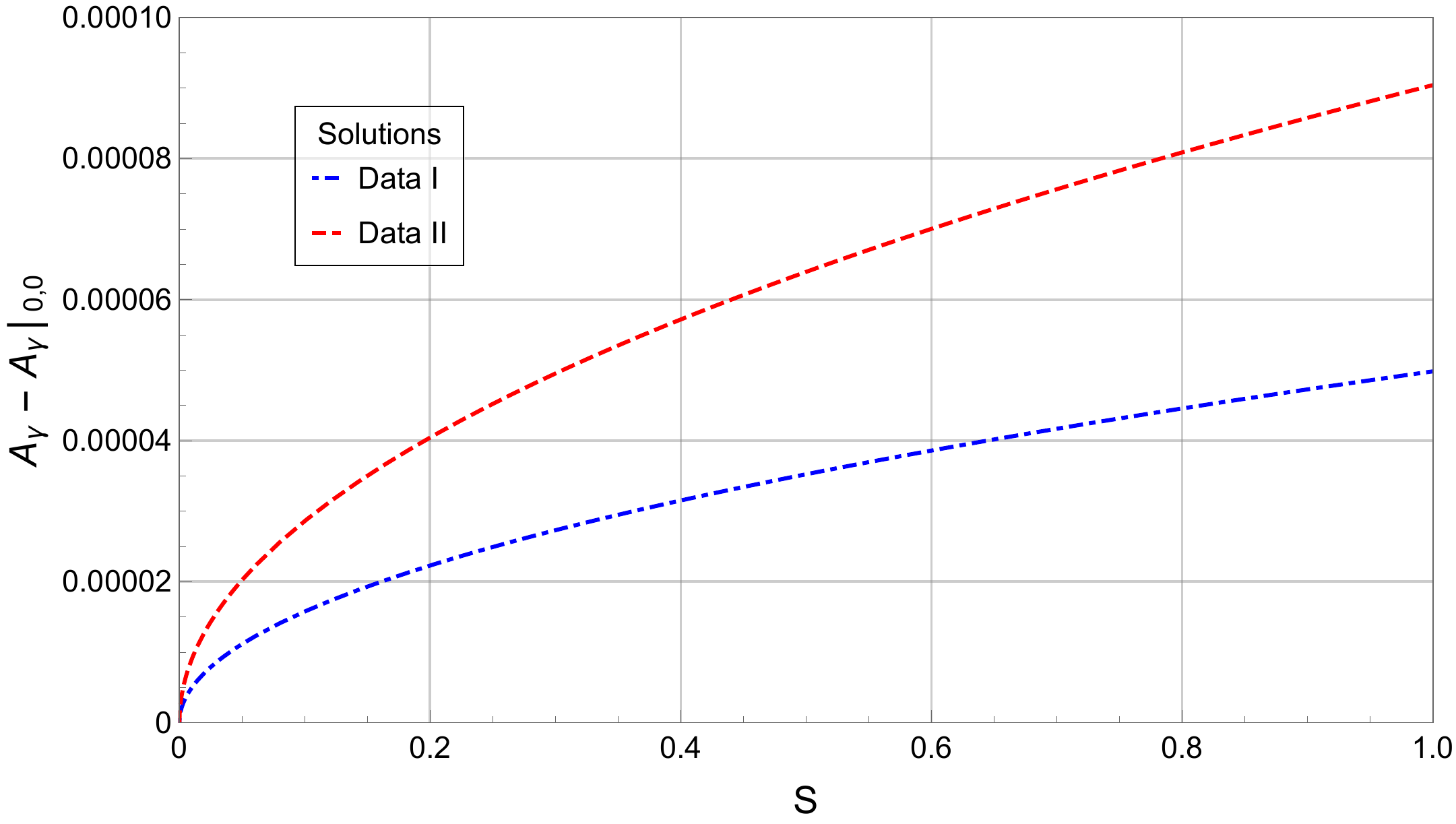}
\caption{Graphical representation of $A_\gamma(S) - A_\gamma|_{0,0}$ from Eq.~(\ref{A_gamma}), where $A_\gamma|_{0,0}$ refers to the values $V_c = 0$ and $a = 0$. For the values: ($V_{cI},a_I$) given by Data I and ($V_{cII},a_{II}$) given by Data II.} 
\label{fig:Agamma_vs_S_model01}
\end{figure}

We introduce two new thermodynamic parameters. The first one refers to $\gamma$, defined as $A_{\gamma} = \frac{\partial M}{\partial \gamma}$, and the second one refers to the critical radius $a$, defined as $A_{a} = \frac{\partial M}{\partial a}$. From Eq.~(\ref{Massa_sol01}) we therefore obtain
\begin{equation}
A_{\gamma}(S,a)=\frac{\sqrt{S}}{2\sqrt{\pi}}\left(a^2+\frac{S}{\pi}\right)^{V^2_c}\label{A_gamma}
\end{equation}
and
\begin{equation}
    A_a(S,\gamma) = \frac{a\gamma V_c^2\sqrt{S}}{\sqrt{\pi}}\left(a^2+\frac{S}{\pi}\right)^{V_c^2-1}\label{Aa_model01} .
\end{equation}
From Eq.~(\ref{A_gamma}) it follows that $A_\gamma \propto \sqrt{S}$, which in conjunction with Fig.~\ref{fig:Massa_vs_S_model01} indicates that more massive BH have larger variations with respect to $\gamma$. This is consistent with Fig.~\ref{fig:Agamma_vs_S_model01}, which shows that $A_\gamma$ has a stronger effect for Data II, where $V_c$, $a$ and $M$ are larger than in Data I. This trend is even more pronounced for high entropy values.

\begin{figure}[htb!]
\includegraphics[width=\linewidth]{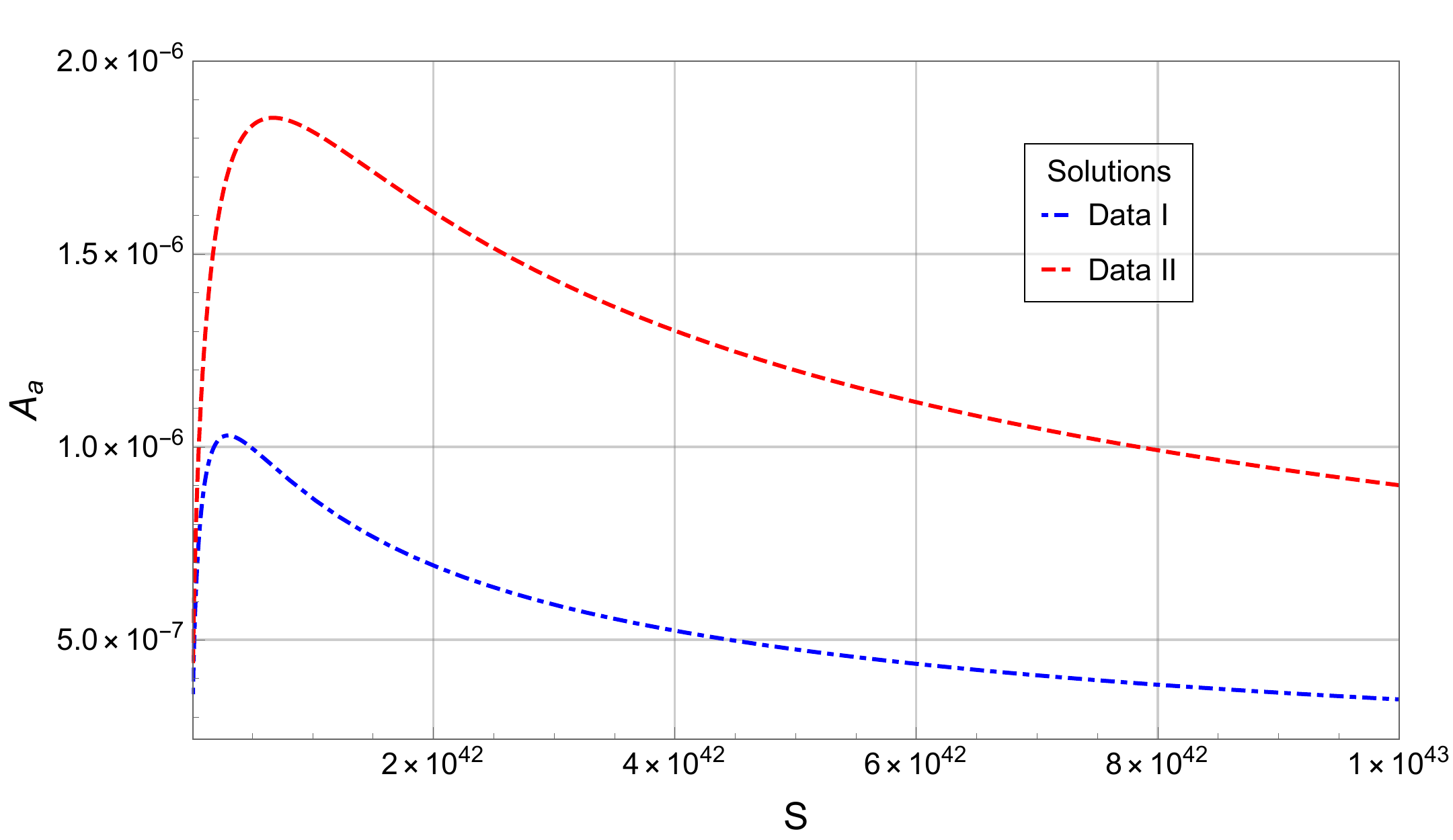}
\caption{Graphical representation of $A_a(S)$ from Eq.~(\ref{Aa_model01}). For the values: ($V_{cI},a_I,\gamma_0$) given by Data I and ($V_{cII},a_{II},\gamma_0$) given by Data II.} 
\label{fig:Aa_vs_S_model01}
\end{figure}

Fig.~\ref{fig:Aa_vs_S_model01}, from Eq.~(\ref{Aa_model01}), shows that systems with larger critical radius have $A_a$ values that are more sensitive to entropy variations, indicating a contribution of the halo to the BH mass. Both curves show that there is an entropy value $S$ that leads to the largest fluctuations of $M$ with respect to the radius $a$. This entropy value is given by $S_m = \frac{a^2\pi}{1 - 2V_c^2}$. This maximum value indicates a transition between different regimes with respect to the significance of the halo parameter $a$ for the BH mass $M$.
\\

% scale=0.24

% scale=0.24

%%%%%%%%%%%%%%%%%%%%%%%%%%%%%%%%%%%%%%%%%%%%%%%%%%%%%%%%%%%%%%%%
\paragraph{Smarr Formula:} 
%%%%%%%%%%%%%%%%%%%%%%%%%%%%%%%%%%%%%%%%%%%%%%%%%%%%%%%%%%%%%%%%

We obtain the Smarr formula by rewriting the mass function, Eq.~(\ref{Massa_sol01}). Using the properties of a homogeneous function \cite{smarr-hankey1972}, we consider the transformation $M(S,\gamma,a)\rightarrow M(q^{c_1}S,q^{c_2}\gamma,q^{c_3}a)$, i.e
\begin{equation}
M=\frac{q^{c_2+c_3+2c_3V^2_c}\gamma \sqrt{S}}{2\sqrt{\pi}}\left(a^2+\frac{S}{\pi}\right)^{V_c^2} ,
\end{equation}
where $c_1=2c_3$. If we define $c_1=1$ and $c_2=-V_c^2$, the mass is represented as a homogeneous function of degree $1/2$,
\begin{equation}
M(q^{c_1}S,q^{c_2}\gamma,q^{c_3}a)=q^{1/2}M(S,\gamma,a) ,
\end{equation}
thus for the Smarr formula,
\begin{equation}
    nM=c_1\frac{\partial M}{\partial S}S+c_2\frac{\partial M}{\partial\gamma}\gamma+c_3\frac{\partial M}{\partial a}a ,
\end{equation}
for which $n = 1/2$. Therefore,
\begin{equation}
    M=2TS-2V_c^2\gamma A_{\gamma}+aA_a\label{eqM} .
\end{equation}
From the mass differential, Eq.~(\ref{eqM}),
\begin{equation}
dM=\frac{\partial M}{\partial S}dS+\frac{\partial M}{\partial \gamma}d\gamma+\frac{\partial M}{\partial a}da  \,.  
\end{equation}
and implies the first law of thermodynamics
\begin{equation}
dM=TdS+A_{\gamma}d\gamma+A_ada .
\end{equation}

%%%%%%%%%%%%%%%%%%%%%%%%%%%%%%%%%%%%%%%%%%%%%%%%%%%%%%%%%%%%%%%%
\paragraph{Surface Gravity:\label{p_grav_sup_model01}} 
%%%%%%%%%%%%%%%%%%%%%%%%%%%%%%%%%%%%%%%%%%%%%%%%%%%%%%%%%%%%%%%%

Just as the entropy is related to the surface of the BH, the temperature reacts directly to the surface gravity $\kappa$ \cite{Bardeen1973},
\begin{equation}
    T=\frac{\kappa}{2\pi} ,\label{T_kappa}
\end{equation}
where
\begin{equation}
    \kappa = \frac{1}{2}\frac{dF(r)}{dr}\Bigg|_{r_h} .\label{kappa_def}
\end{equation}
Since we know $r_h$ from Eq.~(\ref{rh_model01_}) for a restricted scenario, we then obtain $\kappa$ from Eq.~(\ref{kappa_def}) as
\begin{equation}
    \kappa = 2a^{-2V^2_c}MV_c^2\left(a^2+\frac{4a^{-4V^2_c}M^2}{\gamma^2}\right)^{V^2_c-1}+\frac{a^{4V_c^2}\gamma^2}{4M} .\label{kappa_sol01}
\end{equation}
According to the zeroth law of thermodynamics \cite{Bardeen1973}, surface gravity is constant for stationary solutions above the event horizon.

\begin{figure}[htb!]
	\includegraphics[width=\linewidth]{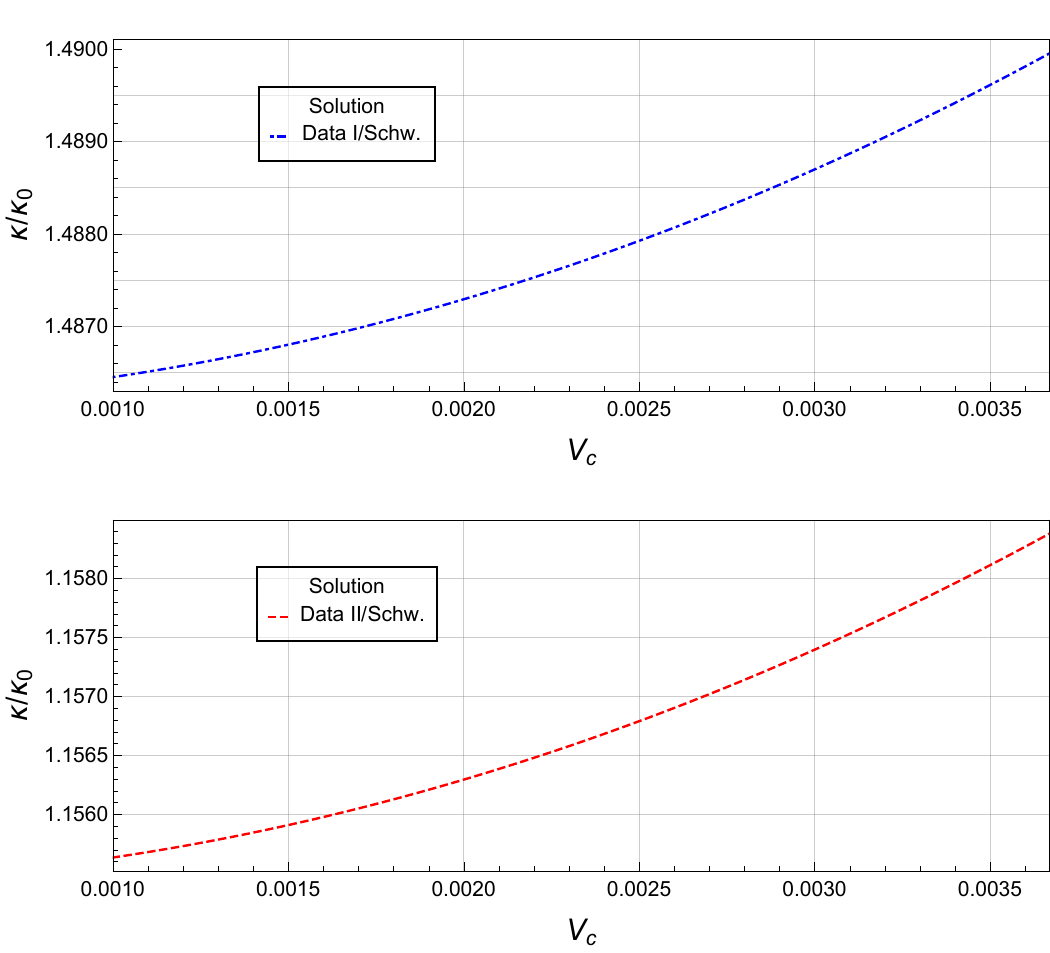}
	\caption{Graphical representation of $\frac{\kappa(V_c)}{\kappa_0}$ for Data I (upper figure) and Data II (lower figure), from Eq.~(\ref{kappa_sol01}), where $\kappa_0=\frac{1}{4M_0}$ refers to the Schwarzschild solution without a halo.} 
	\label{fig:grav_superf_vs_Vc_model01}
\end{figure}

In view of this, the values for the surface gravity $\kappa$ for Data I, Data II and without halo are: $\kappa_I= 5.85313\times10^{-14}$, $\kappa_{II}= 4.55179\times10^{-14}$ and $\kappa_0= 4.76378\times10^{-14}$. 

Figure~\ref{fig:grav_superf_vs_Vc_model01} depicts the relationship between $\frac{\kappa(V_c)}{\kappa_0}$ according to Eq.~(\ref{kappa_sol01}). The behavior of this ratio shows a slight increase due to the increase in velocity, which indicates that the halo affects the gravitational force exerted on the horizon, so that $\kappa \propto V_c$. A larger increase in $\kappa$ occurs when $M$ and $a$ are assigned higher intensities, as can be seen from the different intervals for Data I and II.

Since $\kappa \propto T$, the temperature reacts in a similar way to $V_c$, in accordance with the third law of thermodynamics for BHs \cite{Bardeen1973}, i.e. the surface gravity should not be zero.

From Eq.~(\ref{kappa_sol01}), if $V_c = 0$, we return to the Schwarzschild solution. However, the other parameters cannot be zero due to simple mathematical implications. On the other hand, Eq.~(\ref{Massa_sol01}) shows that if $S \rightarrow 0$, then $M \rightarrow 0$, which in turn can imply a vanishing $\kappa$. However, before $S$ disappears, the black hole reaches a stage where its size is of the same order of magnitude as that of the particles.

As proposed in \cite{Harada2006}, there exists a threshold mass below which a black hole becomes unstable or non-classical due to dominant Hawking evaporation effects \cite{3-Hawking1975}. This mass scale may define a separate class of solutions associated with primordial BHs \cite{HAW-0011, Carr-2023PBH}, which lie beyond the scope of this work.

%%%%%%%%%%%%%%%%%%%%%%%%%%%%%%%%%%%%%%%%%%%%%%%%%%%%%%%%%%%%%%%%
\subsection{Second Solution}
%%%%%%%%%%%%%%%%%%%%%%%%%%%%%%%%%%%%%%%%%%%%%%%%%%%%%%%%%%%%%%%%

%%%%%%%%%%%%%%%%%%%%%%%%%%%%%%%%%%%%%%%%%%%%%%%%%%%%%%%%%%%%%%%%
\subsubsection{Event Horizon}
%%%%%%%%%%%%%%%%%%%%%%%%%%%%%%%%%%%%%%%%%%%%%%%%%%%%%%%%%%%%%%%%

The event horizon radius $r_h$ is determined by solving the equation $F(r_h) = 0$, where
\begin{equation}
	F(r) = 1 - \frac{2M}{r} + \frac{2V_c^2 r^2}{a^2 + r^2}, \label{F(r)_model02}
\end{equation}
as defined with respect to the metric in Eq.~(\ref{metrica_completa_model02}). The corresponding solution for $r_h$ is therefore:
{\small\begin{eqnarray}
	r_h&=&\frac{1}{2
		V_c^2+1}\Bigg[0.264567 \Big(\sqrt{A}
	+216 a^2 M V_c^4+180 a^2 M V_c^2
		\nonumber \\
	&& \hspace{-0.25cm}
	+36 a^2 M+16 M^3\Big)^{1/3}+\left(1.67989 M^2-a^2\sqrt[3]{2} \left(2 
	V_c^2+1\right)\right)
	\nonumber
	\\
	&& \hspace{-0.5cm}
	\times \Bigg(\sqrt{A}+216
	a^2 M V_c^4+180 a^2 M V_c^2+36 a^2 M+16 M^3\Bigg)^{-1/3}
	\nonumber
	\\
	&&
	+\frac{2 M}{3}\Bigg] ,
\end{eqnarray}}
where $A=\left[a^2 M \left(216 V_c^4+180
   V_c^2+36\right)+16 M^3\right]^2+108 \left[a^2 \left(2
   V_c^2+1\right)-\frac{4 M^2}{3}\right]^3$. In the case that $a=V_c=0$, we obtain $r_h=r_{h}^{(\textnormal{sch})}=2M$. 
   
\begin{figure}[tb!]
\includegraphics[scale=0.245]{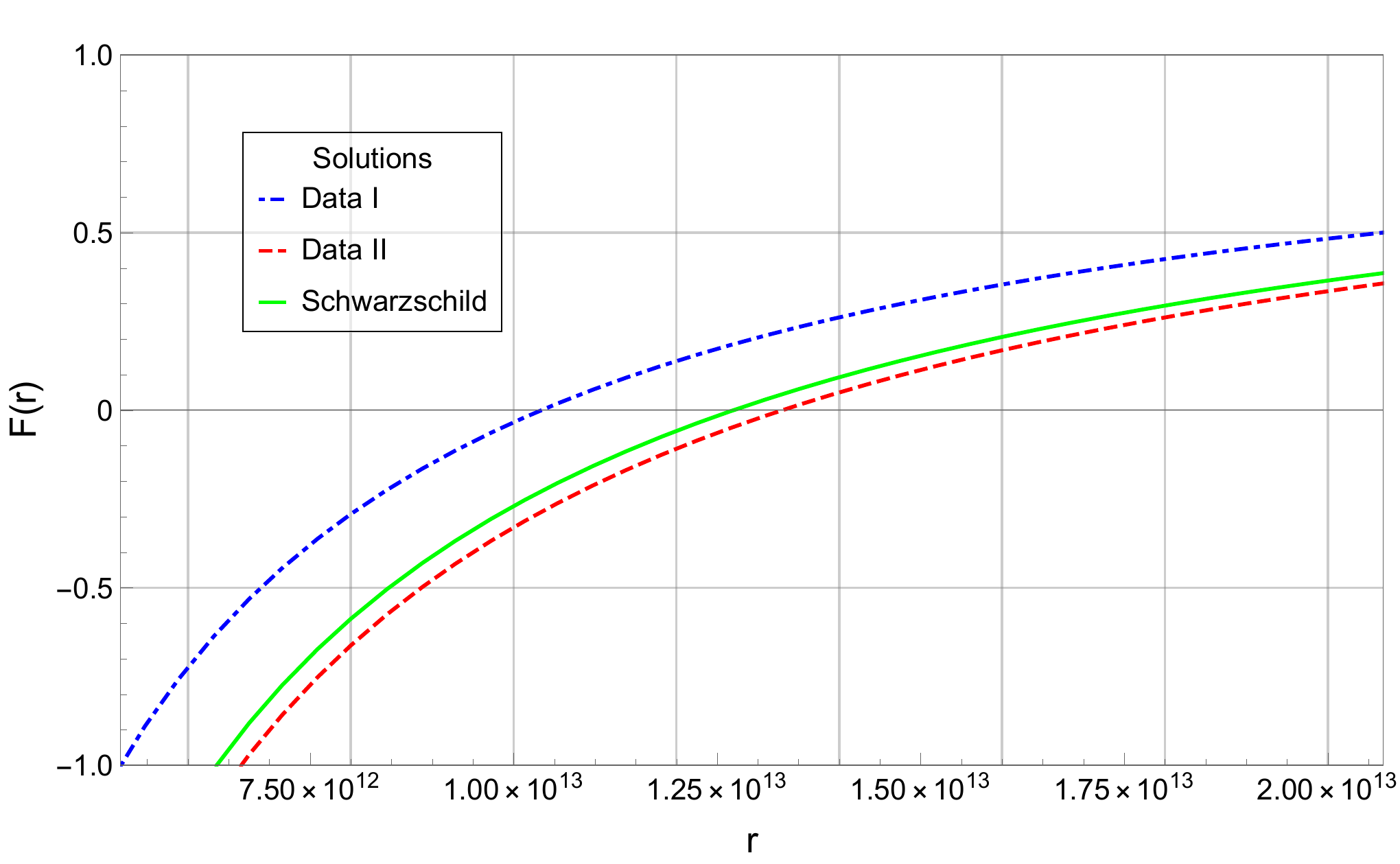}
\caption{Graphical representation of $F(r)$ from Eq.~(\ref{F(r)_model02}), for the following values: ($V_{cI},a_I,M_I$) given by Data I, ($V_{cII},a_{II},M_{II}$) given by Data II, and for the Schwarzschild solution without halo (0,0,$M_0$).} 
\label{fig:F(r)_vs_r_model02}
\end{figure}

Figure~\ref{fig:F(r)_vs_r_model02} shows the behavior of the metric coefficient from Eq.~(\ref{metrica_completa_model02}) and reveals the existence of an event horizon. For Data I, II and without halo, we obtain $r_{hI}=1.034\times10^{13}\textnormal{m}$, $r_{hII}=1.33\times10^{13}\textnormal{m}$ and $r^{(\textnormal{sch})}_{h}=1.27\times10^{13}\textnormal{m}$, respectively. This solution shows that the presence of a DM halo influences the radius of the event horizon. The increase and decrease of the radius, as seen in $r_{hII}>r_h^{(\textnormal{sch})}>r_{hI}$, correspond directly to the mass values $M_{II}>M_0>M_I$, which in turn are consequences of the halo surrounding the BH.
%scale=0.245

%%%%%%%%%%%%%%%%%%%%%%%%%%%%%%%%%%%%%%%%%%%%%%%%%%%%%%%%%%%%%%%%
\subsubsection{Kretschmann Scalar} 
%%%%%%%%%%%%%%%%%%%%%%%%%%%%%%%%%%%%%%%%%%%%%%%%%%%%%%%%%%%%%%%%

To compute the scalar $K$, we use Eq.~(\ref{K_escalar_01}). In this context, the function $f(r)$ is replaced by $F(r)$, as defined in the metric given by Eq.~(\ref{metrica_completa_model02}). Thus,
\begin{eqnarray}
	K &=& \frac{16}{r^6 \left(a^2+r^2\right)^6}
	\Big[3a^{10}M^2\left(a^2+6r^2\right)
		\nonumber \\
	&& +3 a^8 r^4 \left(15
	M^2+2  M r
	V_c^2+2 r^2 V_c^4\right)
		\nonumber \\
	&&
	+2 a^6 r^6 \Big(30  M^2+8 M r V_c^2 +3 r^2
	V_c^4\Big)
	\nonumber \\
	&&
	+a^4 r^8 \left(45
	M^2+12  M r
	V_c^2+19 r^2 V_c^4\right) +2 a^2 r^{10} \times
	\nonumber \\
	&& \hspace{-0.8cm}
	 \times \left(9 M^2+2 r^2 V_c^4\right) +r^{12}
	\Big(3  M^2-2  M r V_c^2 +r^2 V_c^4\Big)\Big] .
	\label{K_sol02_}
\end{eqnarray}

Asymptotically, the curvature follows the same behavior as the first solution:
\begin{equation}
    \lim_{r \to \infty} K\rightarrow 0\,, \qquad {\rm and} \qquad   \lim_{r \to 0} K\rightarrow \infty \,,
\end{equation}
which characterizes flat spacetime as $r\to\infty$ and the singularity at the center of the BH. 

\begin{figure}[htb!]
\includegraphics[scale=0.25]{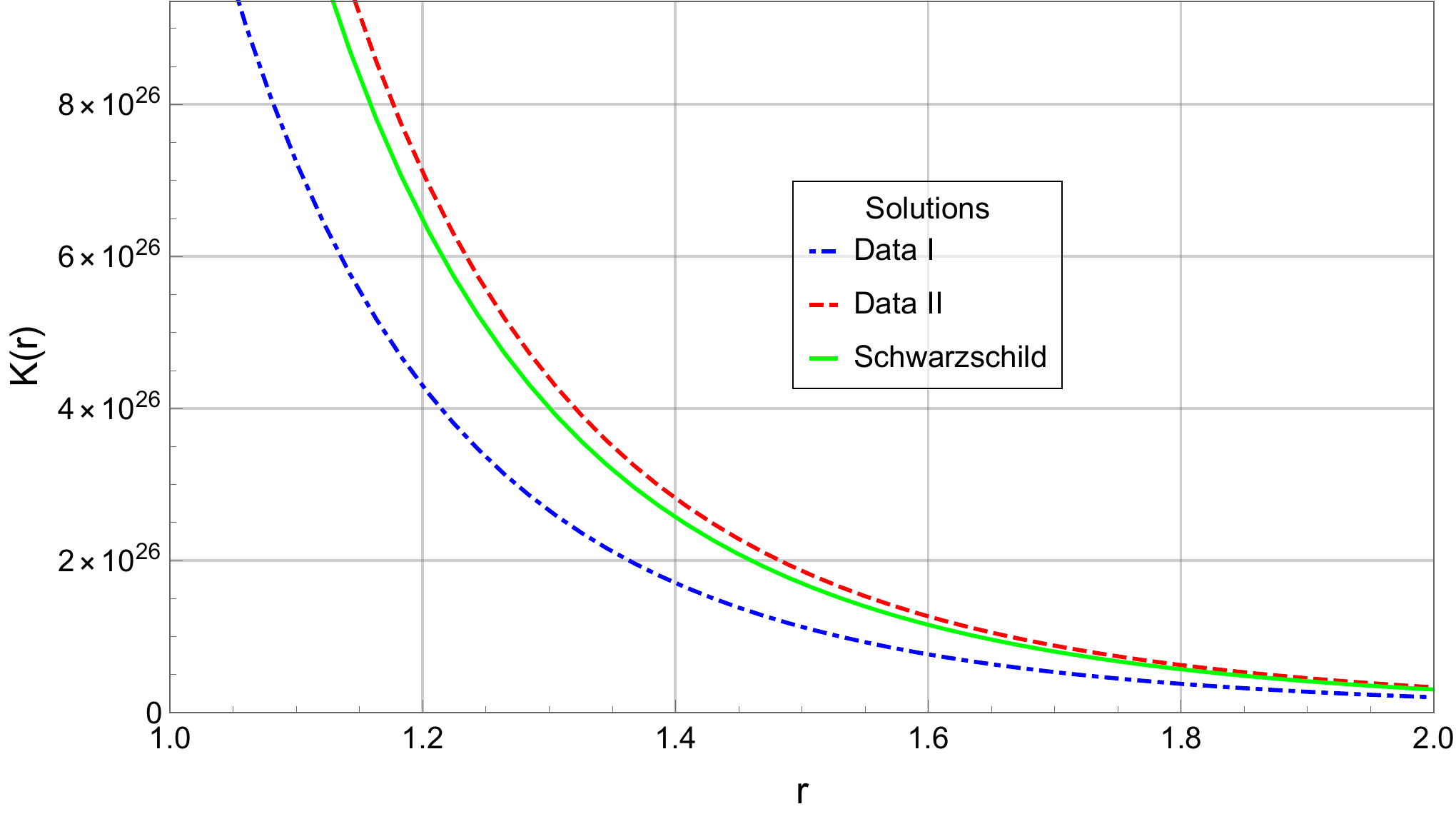}
\caption{Graphical representation of the scalar $K(r)$ from Eq.~(\ref{K_sol02_}) for the values: ($V_{cI}, a_I, M_I$) given by Data I, ($V_{cII}, a_{II}, M_{II}$) given by Data II and for the Schwarzschild solution without a halo (0,0,$M_0$).} 
\label{fig:K(r)_vs_r_model02}
\end{figure}

The most pronounced behavior of $K$ with respect to $r$ is depicted in Fig.~\ref{fig:K(r)_vs_r_model02}.
Similar to the first model, the plot shows that the scalar $K$ is shifted by $V_c$ and $a$, indicating that the curvature intensity around the BH is altered by the halo. The interference in the attenuation of the curvature is more pronounced when $V_c$ and $a$ are larger. These two parameters, as shown in \cite{mod1}, influence the increase in $M$.

%%%%%%%%%%%%%%%%%%%%%%%%%%%%%%%%%%%%%%%%%%%%%%%%%%%%%%%%%%%%%%%%
\subsubsection{Black Hole Shadow Radius}
%%%%%%%%%%%%%%%%%%%%%%%%%%%%%%%%%%%%%%%%%%%%%%%%%%%%%%%%%%%%%%%%

To obtain $r_{ph}$, given the metric in Eq.~(\ref{metrica_completa_model02}), we need to solve Eq.~(\ref{r_ph__metodo}), namely,
\begin{equation}
    1-\frac{2M}{r} + \frac{2V_c^2r^2}{a^2+r^2}=0\label{rph_M02}.
\end{equation}
In this scenario, considering $r \ll a$, we have that
\begin{equation}
r_{ph}\approx3M\label{r_ph_model02} ,
\end{equation}
which shows the independence of the velocity $V_c$. This indicates an influence on $r_{ph}$ by the increase in the mass of the black hole, so that $r_{phII} > r_{ph}^{(\textnormal{halo})}$.

We calculated the photon sphere radius numerically $r_{ph,n}$ without algebraic approximations from Eq.~(\ref{rph_M02}). We obtained the ratio $1 - \frac{r_{ph}}{r_{ph,n}} = 2.39 \times 10^{-35}$ for Data I and $1 - \frac{r_{ph}}{r_{ph,n}} = 2.33 \times 10^{-35}$ for Data II. This shows the validity of Eq.~(\ref{r_ph_model02}).

\begin{figure}[htb!]
\includegraphics[scale=0.25]{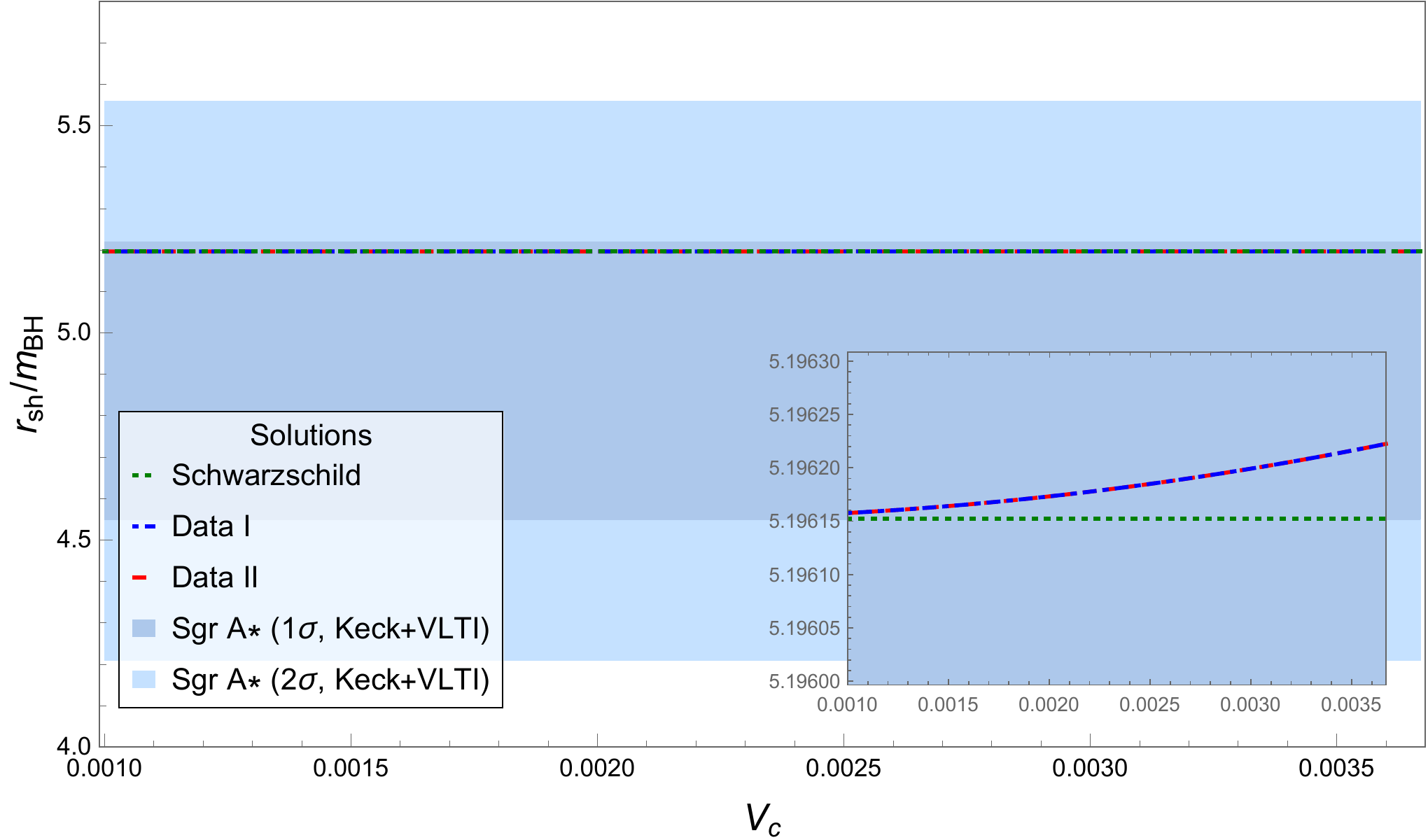}
\caption{Graphical representation of the ratio $r_{sh}/M$, from Eq.~(\ref{rsh_model02f}), with values of $V_c$ within the range of measurements corresponding to $1\sigma$ and $2\sigma$. For the values: ($a_I,M_I$) given by Data I, ($a_{II},M_{II}$) given by Data II and (0,$M_0$) for the SS-type solution without halo.
} 
\label{fig:rshadow_vs_velocidade_model02}
\end{figure}

Using Eq.~(\ref{r_ph_model02}) in Eq.~(\ref{r_sh_GERAL_naoplana}) we obtain the shadow radius as
\begin{equation}
r_{sh}=3M\sqrt{F(r_O)}\left[\frac{1}{3} + \frac{18M^2V^2_c}{\left(a^2 + 9M^2\right)}\right]^{-1/2}\label{rsh_model02f} .
\end{equation}
Figure~\ref{fig:rshadow_vs_velocidade_model02} shows the ratio $r_{sh}/M$ depending on the speed range described by \cite{mod1}, see Tab.~\ref{tab:parametros}. The slight variation of the shadow is a reaction to the high values of $a$ and the low values of $V_c$.

The subfigure in Fig.~\ref{fig:rshadow_vs_velocidade_model02} shows that the shadow of the BH becomes larger as the halo velocity increases. This reflects the idea that the curvature in the region around the BH is also affected by the velocity, not only by the increase in mass of the object.
% scale=0.25

%%%%%%%%%%%%%%%%%%%%%%%%%%%%%%%%%%%%%%%%%%%%%%%%%%%%%%%%%%%%%%%%
\subsubsection{Thermodynamics}
%%%%%%%%%%%%%%%%%%%%%%%%%%%%%%%%%%%%%%%%%%%%%%%%%%%%%%%%%%%%%%%%

\begin{figure}[htb!]
\includegraphics[scale=0.25]{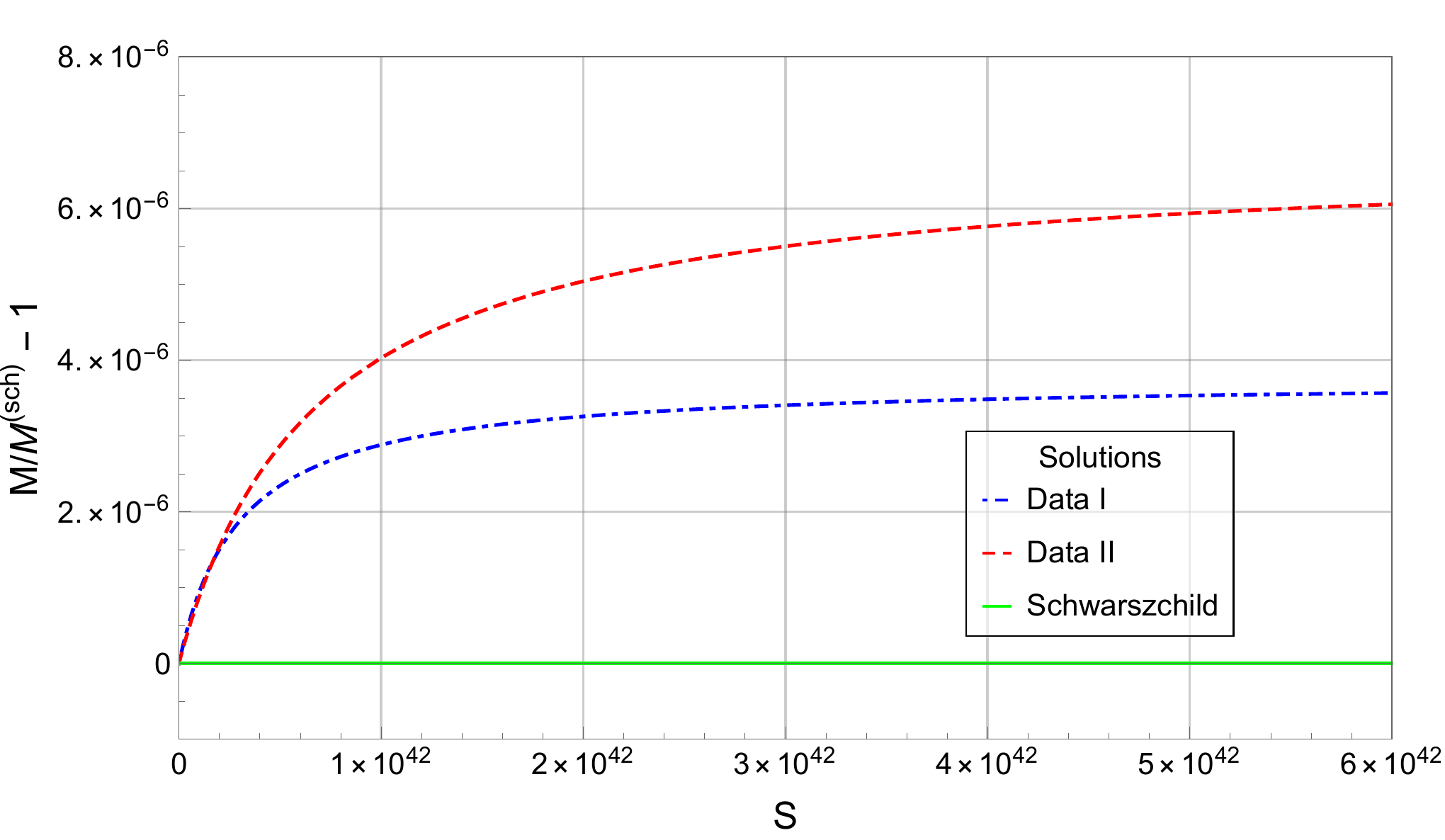}
\caption{Graphical representation of $M(S)/M^{(\textnormal{sch})} -1$, from Eq.~(\ref{Massa_sol02}). For the parameter sets: ($V_{cI},a_I$), given by Data I, ($V_{cII},a_{II}$), given by Data II, and the Schwarzschild solution without a halo (0,0).} 
\label{fig:Massa_vs_S_model02}
\end{figure}

Using the same relationship between the entropy and the radius of the event horizon from Eq.~(\ref{S}), we substitute $r_h$ into $F(r_h)=0$ from Eq.~(\ref{F(r)_model02}) and solve to obtain the following expression for the BH mass,
\begin{equation}
   M(S,V_c,a) = \frac{\sqrt{S}\left(a^2\pi+S+2SV_c^2\right)}{2\sqrt{\pi}\left(a^2\pi+S\right)}\label{Massa_sol02} \ .
\end{equation}
Figure~\ref{fig:Massa_vs_S_model02}, from Eq.~(\ref{Massa_sol02}), shows an increase in mass due to the increase in entropy. This effect is more pronounced for higher values of $(V_c,a)$. For large entropy values, the expression becomes constant,
\begin{equation}
    \lim_{S\to \infty}\left[\frac{M}{M^{(sch)}}-1\right]=2V_c^2 \ ,
\end{equation}
whereby a higher increase in mass is emphasized for Data II, since $V_{cII}>V_{cI}$.
% scale=0.25

To obtain the temperature, we use $T=\frac{\partial M}{\partial S}$, from Eq.~(\ref{Massa_sol02}). Thus,
\begin{equation}
T(S,V_c,a)=\frac{a^4\pi^2+S^2\left(1+2V^2_c\right)+2a^2\pi S\left(1+3V^2_c\right)}{4\sqrt{\pi S}\left(a^2\pi+S\right)^2}. \label{temperatura_sol02}
\end{equation}

\begin{figure}[htb!]
\includegraphics[scale=0.235]{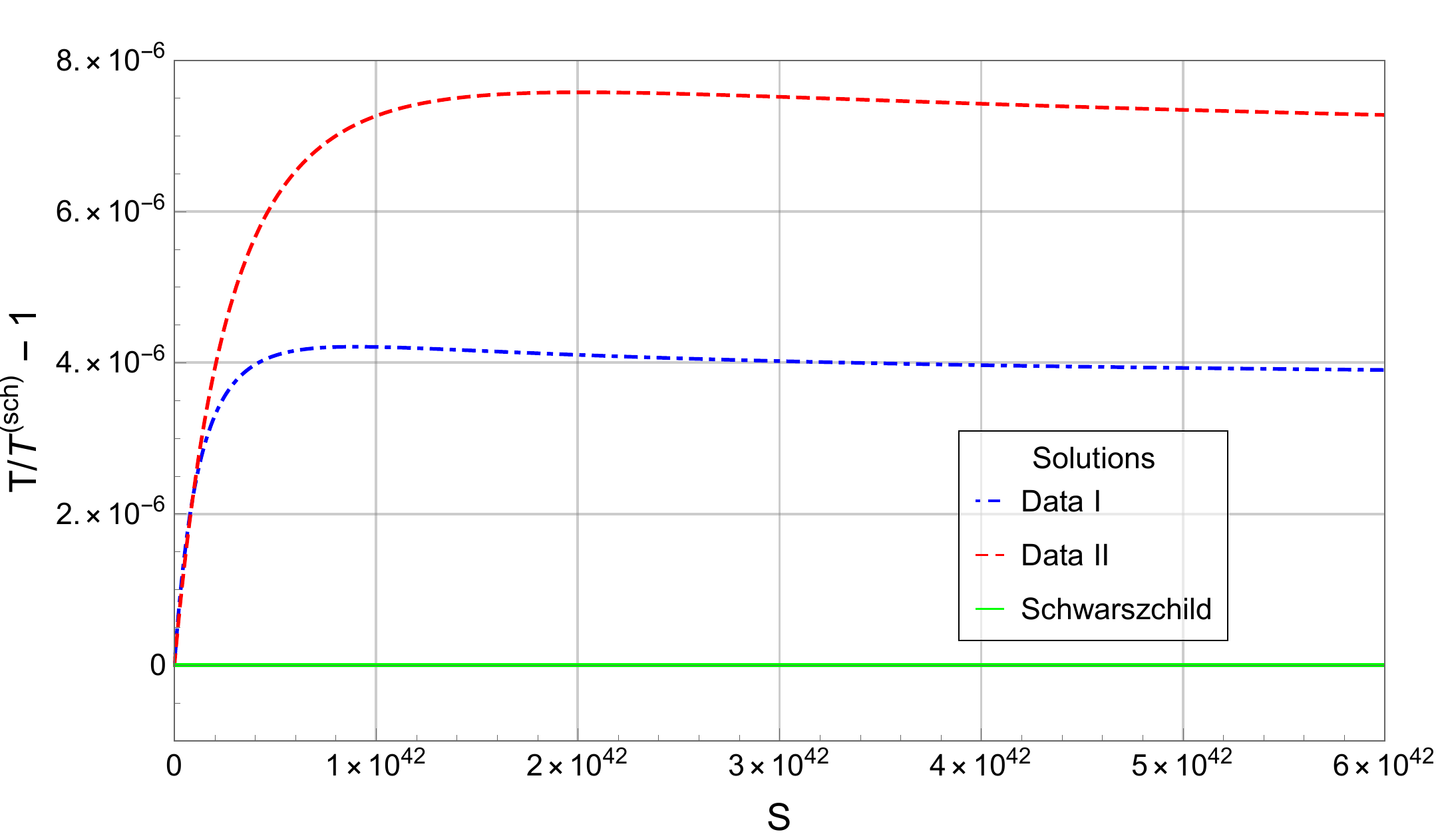}
\caption{Graphical representation of $T(S)/T^{(sch)} - 1$, from Eq.~(\ref{temperatura_sol02}). For the values: ($V_{cI},a_I$) given by Data I, and ($V_{cII},a_{II}$) given by Data II, and for the Schwarzschild solution without a halo (0,0).} 
\label{fig:T_vs_S_model02}
\end{figure}

Since the halo influences the mass $M(S)$, a similar behavior is expected for the temperature $T(S)$. Such behavior can be seen in Fig.~\ref{fig:T_vs_S_model02} from Eq.~(\ref{temperatura_sol02}), which indicates a certain increase in the surface temperature of the BH due to high values of entropy. The upper limit of this variation is
\begin{equation}
    \lim_{S\to\infty}\left[\frac{T}{T^{(sch)}}-1\right]=2V^2_c ,
\end{equation}
where a direct dependence on the speed can also be observed.

Similar to the first solution, in this second model we define two new parameters related to the velocity $V_c$, namely $A_{V_c}=\frac{\partial M}{\partial V_c}$, and to the radius $a$, namely $A_a=\frac{\partial M}{\partial a}$. From Eq.~(\ref{Massa_sol02}) we therefore obtain
\begin{equation}
    A_{V_c}(S,a)=\frac{2S^{3/2}V_c}{\sqrt{\pi}\left(a^2\pi+S\right)}\,,
    \label{Avc_model02}
\end{equation}
and
\begin{equation}
    A_a(S,V_c)=-\frac{2a\sqrt{\pi }S^{3/2}V^2_c}{\left(a^2\pi+S\right)^2}\label{Aa_model02_} .
\end{equation}

\begin{figure}[htb!]
\includegraphics[scale=0.24]{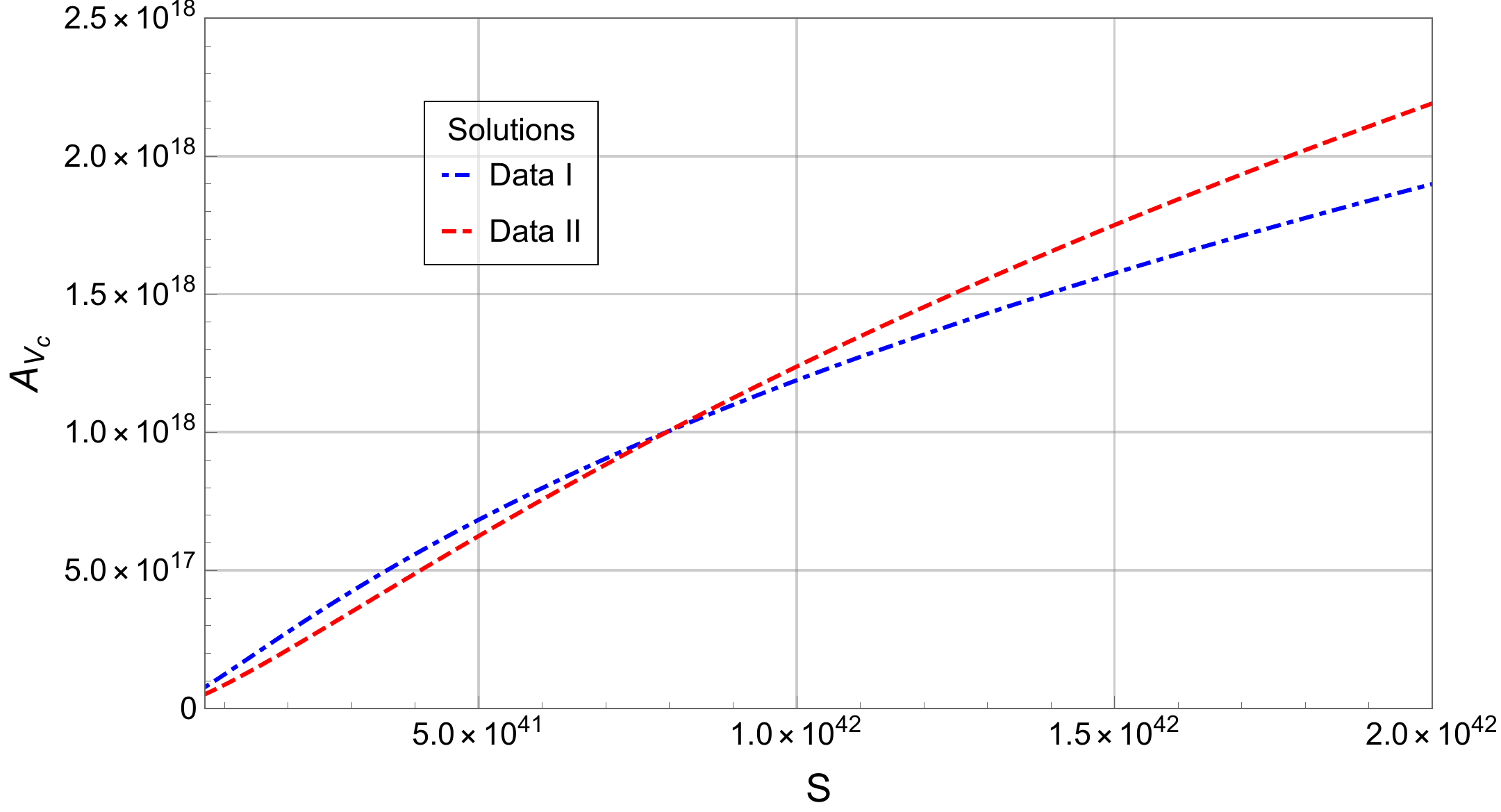}
\caption{Graphical representation of $A_{V_c}(S)$, from Eq.~(\ref{Avc_model02}). For values: ($V_{cI},a_I$) given by Data I and ($V_{cII},a_{II}$) given by Data II.} 
\label{fig:Av_vs_S_model02}
\end{figure}

Figure~\ref{fig:Av_vs_S_model02}, from Eq.~(\ref{Avc_model02}), shows the influence of $V_c$ on $M$ through $A_{V_c}(S)$. The BH mass reacts significantly to changes in the halo velocity as the entropy increases. The exchange of the order of the graphical lines occurs when $S \sim \pi a^2$ marks a transition of specific influence between Data I and II. Before this transition area, the velocity has a greater influence on the mass when the parameters are less intense, in contrast to the scenario after this transition.

\begin{figure}[htb!]
\includegraphics[scale=0.24]{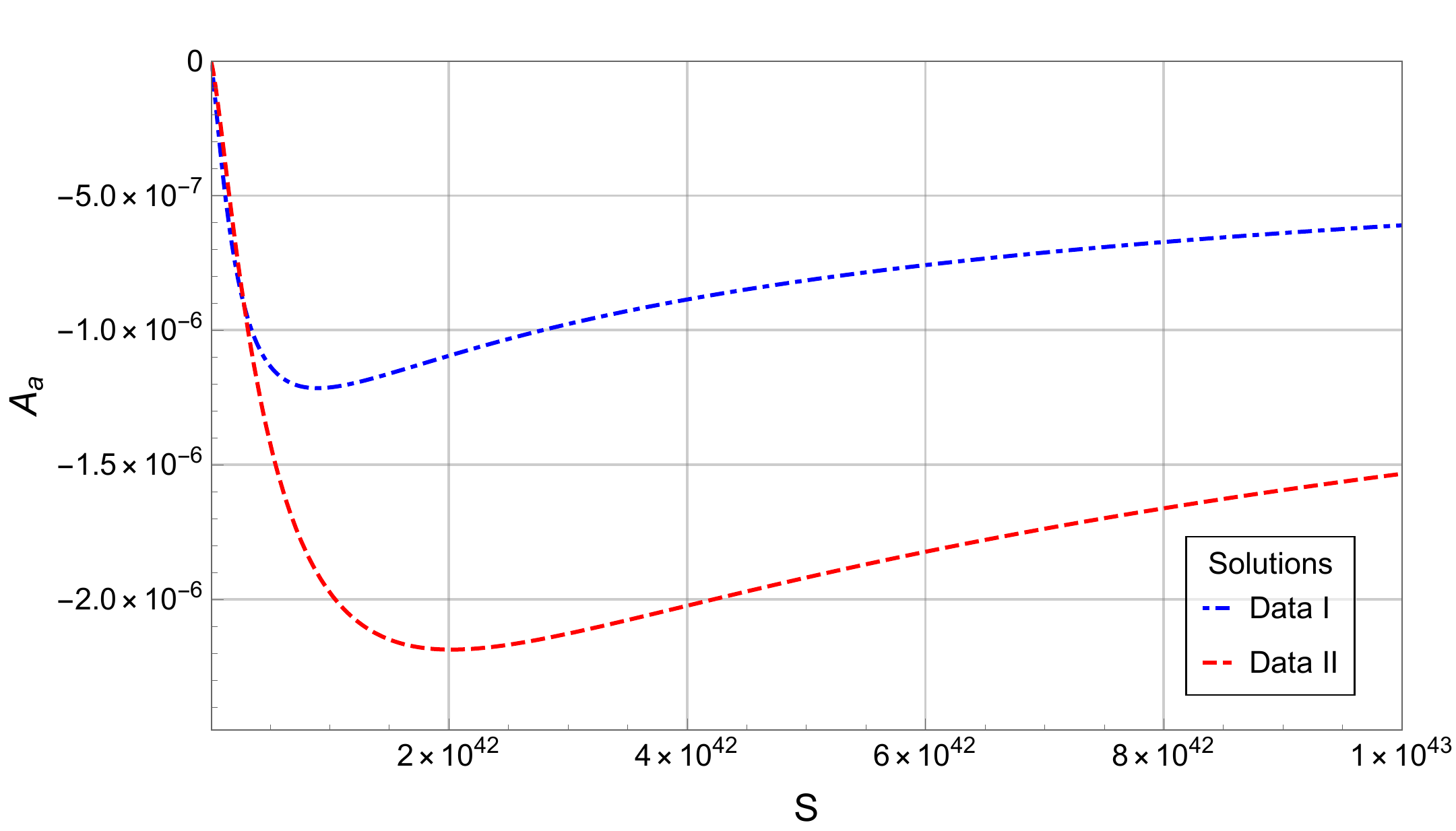}
\caption{Graphical representation of $A_a(S)$, from Eq.~(\ref{Aa_model02_}). For values: ($V_{cI},a_I$) given by Data I and ($V_{cII},a_{II}$) given by Data II.} 
\label{fig:Aa_vs_S_model02}
\end{figure}

Figure~\ref{fig:Aa_vs_S_model02}, from Eq.~(\ref{Aa_model02_}), represents $A_a(S)$, which measures the response of mass with respect to radius in relation to the variation of entropy. It is less attenuated for halos with higher $V_c$ and $a$. Nevertheless, both sets of values show a maximum influence point for $A_{a'}$ at entropy $S' = 3\pi a^2$, independent of the radius, with $A_{a'} = -\frac{3\sqrt{3}V_c^2}{8}$.

In the range of $S$, with the exception of $S = 0$, from Eq.~(\ref{Aa_model02_}), we can approximately determine the relationship $\frac{\delta M}{\delta a} < 0$. Therefore, $\delta M > 0 \to \delta a < 0$, since $M^{(\textnormal{halo})} > M_{0}$. The term $\delta a$ indicates the change in radius due to a certain entropy value, which leads us to the realization that the halo decreases with increasing mass. The intensity of $|\delta a|$ increases up to $S = S'$ and then gradually decreases. This implies the classification of two regimes around the stable point $S'$, since $\left(\frac{\partial^2 A_a}{\partial S^2}\right)_{S'} > 0$.

% scale=0.235
% scale=0.24
% scale=0.24
%%%%%%%%%%%%%%%%%%%%%%%%%%%%%%%%%%%%%%%%%%%%%%%%%%%%%%%%%%%%%%%%%%%%%%%%%%%%%%%%%%%%%%%%%%%%%%%%%%%%%%%%%%%%%%%%%

\paragraph{Smarr Formula:}
To define the Smarr formula, we rewrite the mass function from Eq.~(\ref{Massa_sol02}) as $M(S,V_c,a)\rightarrow M(q^{c_1}S,q^{c_2}V_c,q^{c_3}a)$, where $c_1=2c_3$. If we set $c_1=1$ and $c_2=0$, we get
\begin{equation}
M(q^{c_1}S,q^{c_2}V_c,q^{c_3}a)=q^{1/2}M(S,V_c,a) \ ,
\end{equation}
which leads to the Smarr formula,
\begin{equation}
    M=2TS+aA_a ,
\end{equation}
which is independent of $A_{V_c}$. The first law of thermodynamics for this solution is therefore
\begin{equation}
 dM=TdS+A_{V_c}dV_c + A_a da \ ,
\end{equation}
where the contributions of the halo to the mass are emphasized by its velocity and its distance from the BH center.
%%%%%%%%%%%%%%%%%%%%%%%%%%%%%%%%%%%%%%%%%%%%%%%%%%%%%%%%%%%%%%%%%%%%%%%%%%%%%%%%%%%%%%%%%%%%%%%%%%%%%%%%%%%%%%%%%%

\paragraph{Surface Gravity:} 
As in the first model, in this solution we use Eq.~(\ref{T_kappa}) to relate the temperature to the surface gravity of the BH with halo. From Eq.~(\ref{kappa_def}) we obtain
\begin{eqnarray}
	\kappa&=&-\frac{2V^2_c(\frac{2M}{3}+B_1+B_2)^3}{(2V^2_c+1)^3[a^2+\frac{1}{(2V^2_c+1)^2}(\frac{2M}{3}+B_1+B_2)^2]^2} 
		\nonumber\\
	&&+ \frac{2V_c^2(\frac{2M}{3}+B_1+B_2)}{(2V_c^2+1)(a^2+\frac{1}{(2V_c^2+1)^2}(\frac{2M}{3}+B_1+B_2)^2)} 
		\nonumber\\
	&&+ \frac{M(2V^2_c+1)^2}{(\frac{2M}{3}+B_1+B_2)^2},
	\label{grav_sup_model02}
\end{eqnarray}
where $B_1=0.2646[36M(6V^4_c+5V^2_c+1)a^2+2(8M^3+A_1)]^{1/3}$, $B_2=\frac{(1.3333M^2-(2V^2_c+1)a^2)}{[8M^3+18a^2(6V^4_c+5V^2_c+1)M+A_1]^{1/3}}$, and $A_1=3a(2V_c^2+1)\sqrt{3[(2V^2_c+1)a^4+4M^2(27V^4_c+18V^2_c+2)a^2+16M^4]}$. The values of $\kappa$ for each data set are $\kappa_I=3.79313\times10^{-16}$, $\kappa_{II}=2.22423\times10^{-16}$, and $\kappa_0=3.93701\times10^{-14}$, respectively for Data I, Data II and for the case without halo.

\begin{figure}[tb!]
\includegraphics[width=\linewidth]{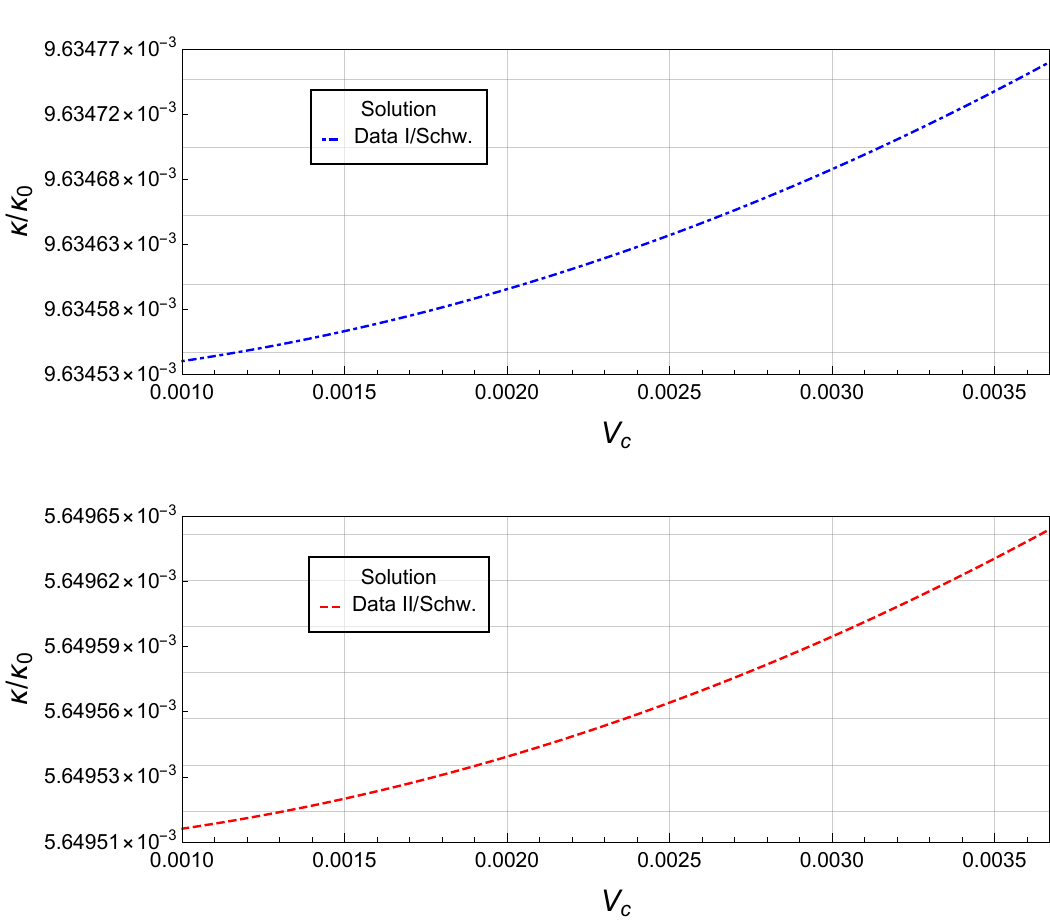}
\caption{Graphical representation of $\frac{\kappa(V_c)}{\kappa_0}$ for Data I (upper figure) and Data II (lower figure), from Eq.~(\ref{grav_sup_model02}), where $\kappa_0=\frac{1}{4M_0}$ refers to the Schwarzschild solution without halo.} 
\label{fig:grav_superf_vs_Vc_model02}
\end{figure}

From Eq.~(\ref{grav_sup_model02}), Fig.~\ref{fig:grav_superf_vs_Vc_model02} shows the ratio $\frac{\kappa_{II}(V_c)}{\kappa_0}$, which shows that the halo causes a slight change in $\kappa$ as a function of $V_c$. Since $T \propto \kappa$, the increase in the halo velocity implies a corresponding increase in the temperature of the event horizon. For $T \rightarrow 0$, which leads to $\kappa \rightarrow 0$, we set $S \rightarrow \infty$, as shown in Fig.~\ref{fig:T_vs_S_model02}. However, this would lead to a BH with zero mass, as shown in Fig.~\ref{fig:Massa_vs_S_model02}. This scenario is inconsistent as it violates the third law of thermodynamics. As with the first model, we emphasize that for $M$ with low order values, quantum effects such as Hawking evaporation \cite{3-Hawking1975} occur, which are not included in the models studied here\footnote{This asymptotic behavior of $\kappa \rightarrow 0$ is repeated in the first model, see \ref{p_grav_sup_model01}, so that the discussion in its last paragraph also applies to this second solution.}.

%width=\linewidth scale=0.475

%%%%%%%%%%%%%%%%%%%%%%%%%%%%%%%%%%%%%%%%%%%%%%%%%%%%%%%%%%%%%%%%
\section{Conclusion}\label{sec:concl}
%%%%%%%%%%%%%%%%%%%%%%%%%%%%%%%%%%%%%%%%%%%%%%%%%%%%%%%%%%%%%%%%

In this work, we proposed two Schwarzschild-type solutions describing a BH embedded in a DM halo. Both models adopt the same DM density profile introduced in \cite{mod1}, characterized by the critical velocity \( V_c \) and the critical radius \( a \) of the halo. 
The DM density profile is constrained by Hubble Space
Telescope (HST) data, stellar dynamics, and globular cluster (GC) measurements of the elliptical
galaxy NGC 4649 (M60).
The first solution, based on \cite{AA01}, introduces a parameter \( \gamma \) that recovers the standard Schwarzschild (SS) solution when \( \gamma = 1 \). The second model, following \cite{DM+materiaUsual}, modifies the metric by including a term proportional to \( V_c^2 \), reverting to the SS case when \( V_c = 0 \). Parameter values were obtained from \cite{mod1}, using two representative data sets — Data I and Data II — to examine the influence of the halo on the BH properties.

For both solutions, a single event horizon exists, with its radius modified by the halo parameters \( M \), \( V_c \), and \( a \). The Kretschmann scalar \( K \) was computed and shown to exhibit standard singular behavior at \( r \rightarrow 0 \) and asymptotic flatness as \( r \rightarrow \infty \). In the intermediate region, the presence of the halo alters the curvature, with the impact of \( M \) — and by extension \( V_c \) and \( a \) — modifying the local geometry. We also investigated the photon sphere radius \( r_{ph} \) and the resulting shadow radius \( r_{sh} \), under the approximations \( r \ll a \) and \( V_c \ll 1 \), to assess potential observational signatures of the DM halo.

In the first model, the shadow radius is indirectly affected by the halo through the parameter \( \gamma \), which varies with the observer’s location. For distant observers (\( r \gg a \)), we find \( \gamma_O \sim r_O^{-2V_c^2} \), while for near-horizon observers (\( r \ll a \)), \( \gamma_c \sim a^{-2V_c^2} \), with \( \gamma_O \lesssim \gamma_c \). This implies a larger shadow radius observed at large distances. In the second model, \( r_{sh} \) increases directly with \( V_c \), indicating that BH shadow measurements may provide a way to detect halo-induced gravitational effects.

Our thermodynamic analysis revealed that the mass–entropy relation approximately satisfies \( M^2 \propto S \) due to the smallness of \( V_c \). In both models, the mass increases with the inclusion of the DM halo, with \( M > M^{(\textnormal{Sch})} \) for higher values of \( V_c \) and \( a \). As entropy \( S \to \infty \), the mass approaches \( M \sim (1 + 2V_c^2)M^{(\textnormal{Sch})} \), highlighting the explicit contribution of the halo to the system's total energy content. The temperature also increases in both models, with Data II showing a more pronounced rise compared to Data I, reflecting a stronger halo influence.

New thermodynamic potentials were introduced to account for the additional parameters. In the first model, potentials \( A_\gamma \) and \( A_a \) describe how variations in \( \gamma \) and \( a \) affect the BH mass. Lower \( \gamma \) enhances \( M \), in line with the increased shadow radius. \( A_a \) captures how the entropy-dependent growth of \( M \) relates to changes in \( a \). In the second model, \( A_{V_c} \) and \( A_a \) govern the response of \( M \) to the halo properties. We found \( A_a < 0 \) across the entropy range, with \( \Delta M > 0 \Rightarrow \Delta a < 0 \), indicating a stability point \( S' \) around which the system exhibits different dynamical behavior, depending on whether \( S < S' \) or \( S > S' \).

Both models yielded Smarr-type relations derived from mass homogeneity. In the first model, the mass was expressed in terms of \( S \), \( \gamma \), and \( a \), enabling the derivation of a consistent first law of thermodynamics. In the second model, the mass was written as a function of \( S \), \( V_c \), and \( a \), leading to a simplified Smarr formula where the \( V_c \) term does not contribute directly. The corresponding first law was derived accordingly. We also discussed violations of the third law in both models, as \( M \rightarrow 0 \) leads to \( \kappa \rightarrow 0 \). However, quantum corrections may allow a restoration of the third law under a more refined framework, including for the limiting case where \( T \rightarrow 0 \).

In summary, the presence of a DM halo induces significant modifications to both the geometric structure and thermodynamic behavior of Schwarzschild-type black holes. The results obtained in this work emphasize the non-negligible role of halo parameters, such as \( V_c \) and \( a \), in shaping observable and intrinsic black hole properties. These findings open new avenues for understanding the astrophysical relevance of dark matter in strong-gravity environments and suggest that black hole observables—such as the shadow radius and thermodynamic quantities—can serve as indirect probes of the dark matter distribution in galactic cores.

Future work will aim to deepen the physical interpretation of these models through a variety of theoretical and observational approaches. One key direction involves the stability analysis of the proposed solutions under linear and nonlinear perturbations. This will clarify whether the halo-modified geometries remain dynamically viable over astrophysical timescales. Another crucial line of research will focus on the study of gravitational lensing in these spacetimes. By examining how the deflection of light is influenced by the halo parameters, we may be able to constrain \( V_c \) and \( a \) through precise lensing observations around SMBHs.

Additionally, we plan to explore the role of quasi-periodic oscillations in accretion disks as a means of inferring the total mass of black holes embedded in DM halos. The modification of orbital frequencies in the presence of the halo could provide a complementary observational signature. Other possible extensions include the investigation of the Hawking radiation spectrum in halo-influenced metrics, the behavior of scalar and electromagnetic fields in these backgrounds, and the potential for testing modified gravity theories within the same framework. Collectively, these directions offer a rich landscape for expanding our theoretical models and bridging the gap with observational astrophysics and cosmology.

%%%%%%%%%%%%%%%%%%%%%%%%%%%%%%%%%%%%%%%%%%%%%%%%%%%%%%%%%%%%%%%%
\acknowledgments{MER thanks Conselho Nacional de Desenvolvimento
Cient\'{\i}fico e Tecnol\'ogico - CNPq, Brazil, for partial financial support. This study was financed in part by the
Coordena\c{c}\~{a}o de Aperfei\c{c}oamento de Pessoal de N\'{\i}vel Superior - Brasil (CAPES) - Finance Code 001. FSNL acknowledges support from the Funda\c{c}\~{a}o para a Ci\^{e}ncia
e a Tecnologia (FCT) Scientific Employment Stimulus contract with reference CEECINST/00032/2018, and
funding through the research grants UIDB/04434/2020,
UIDP/04434/2020 and PTDC/FIS-AST/0054/2021.}

%%%%%%%%%%%%%%%%%%%%%%%%%%%%%%%%%%%%%%%%%%%%%%%%%%%%%%%%%%%%%%%%

%%%%%%%%%%%%%%%%%%%%%%%%%%%%%%%%%%%%%%%%%%%%%%%%%%%%%%%%%%%%%%%%

%%%%%%%%%%%%%%%%%%%%%%%%%%%%%%%%%%%%%%%%%%%%%%%%%%%%%%%%%%%%%%%%

%%%%%%%%%%%%%%%%%%%%%%%%%%%%%%%%%%%%%%%%%%%%%%%%%%%%%%%%%%%%%%%%
\end{document}